\documentclass[conference]{IEEEtran}

\usepackage{subcaption}
\usepackage{tabularx}
\usepackage{booktabs}
\usepackage{tikz}
\usetikzlibrary{calc, arrows, arrows.meta}
\usepackage[siunitx]{circuitikz}
\usepackage{pgfplots}
\usepgfplotslibrary{groupplots}
\pgfplotsset{compat=1.7}
\usepackage{amsfonts}
\usepackage[pdfa]{hyperref}
\usepackage{comment}
\usepackage{pifont}
\usepackage{enumitem}
\usepackage{amsmath}
\usepackage{float}
\usepackage{flushend}
\usepackage[final]{pdfpages}

\newcommand\thefontsize{The current font size is: \f@size pt}

\newcommand{\etal}{\textit{et al}. }

\newcommand{\tabhead}{\textbf}

\newif\ifrev
\revtrue
\ifrev
  \newcommand{\yier}[1]{{\color{blue} [Yier: #1]}}
  \newcommand{\haoqi}[1]{{\color{red} [Haoqi: #1]}}
  \newcommand{\shuo}[1]{{\color{green} [Shuo: #1]}}
  \newcommand{\dean}[1]{{\color{cyan} [Dean: #1]}}
\else
  \newcommand{\directions}[1]{}
  \newcommand{\haoqi}[1]{}
  \newcommand{\yier}[1]{}
  \newcommand{\shuo}[1]{}
  \newcommand{\dean}[1]
\fi

\begin{document}
\title{Invisible Finger: Practical Electromagnetic Interference Attack on Touchscreen-based Electronic Devices} 

\author{%
 \IEEEauthorblockN{%
    Haoqi Shan\IEEEauthorrefmark{1}\textsuperscript{\textsection}, 
    Boyi Zhang\IEEEauthorrefmark{1}\textsuperscript{\textsection}, 
    Zihao Zhan\IEEEauthorrefmark{1},
    Dean Sullivan\IEEEauthorrefmark{2}, 
    Shuo Wang\IEEEauthorrefmark{1},
    Yier Jin\IEEEauthorrefmark{1}%
    }%
    \IEEEauthorblockA{\IEEEauthorrefmark{1}University of Florida \\\{haoqi.shan, zby0070, zhan.zihao\}@ufl.edu, \{shuo.wang, yier.jin\}@ece.ufl.edu}
    \IEEEauthorblockA{\IEEEauthorrefmark{2}University of New Hampshire
    \\\{dean.sullivan\}@unh.edu}
}

\maketitle

\begin{NoHyper}
\begingroup\renewcommand\thefootnote{\textsection}
\footnotetext{These two authors contribute equally to the work.}
\endgroup
\end{NoHyper}

\begin{abstract}
Touchscreen-based electronic devices such as smart phones and smart tablets are
widely used in our daily life. While the security of electronic devices have
been heavily investigated recently, the resilience of touchscreens against
various attacks has yet to be thoroughly investigated. In this paper, for the
first time, we show that touchscreen-based electronic devices are vulnerable to
intentional electromagnetic interference (IEMI) attacks in a systematic way and
how to conduct this attack in a practical way. Our contribution lies in not just
demonstrating the attack, but also analyzing and quantifying the underlying
mechanism allowing the novel IEMI attack on touchscreens in detail. We show how
to calculate both the minimum amount of electric field and signal frequency
required to induce touchscreen ghost touches. We further analyze our IEMI attack
on real touchscreens with different magnitudes, frequencies, duration, and
multitouch patterns. The mechanism of controlling the touchscreen-enabled
electronic devices with IEMI signals is also elaborated. We design and evaluate an
out-of-sight touchscreen locator and touch injection feedback mechanism to
assist a practical IEMI attack. Our attack works directly on the touchscreen
circuit regardless of the touchscreen scanning mechanism or operating system.
Our attack can inject short-tap, long-press, and omni-directional gestures on
touchscreens from a distance larger than the average thickness of common
tabletops. Compared with the state-of-the-art touchscreen attack, ours
can accurately inject different types of touch events without the need for sensing
signal synchronization, which makes our attack more robust and practical. In
addition, rather than showing a simple proof-of-concept attack, we present and
demonstrate the first ready-to-use IEMI based touchscreen attack vector with
end-to-end attack scenarios. 
\end{abstract}

\section{Introduction} \label{sec:intro}

Consumer electronic devices with touchscreens, such as smartphones, tablets, and
laptops, have become integral parts of our daily lives because touchscreen technology is both convenient and intuitive to use. In practice, touchscreens recognize a touch event by sensing the electric field of the electrodes under the screen, thereby allowing people to give commands by performing touch, swipe, and other gestures. 
The commands are then converted to electric signals and help control the systems/apps in the target device. For vehicles or medical devices incorporating touchscreens, their correct functionality is tied to user safety.

\begin{figure}[t!]
\centering
\includegraphics[width=0.98\columnwidth]{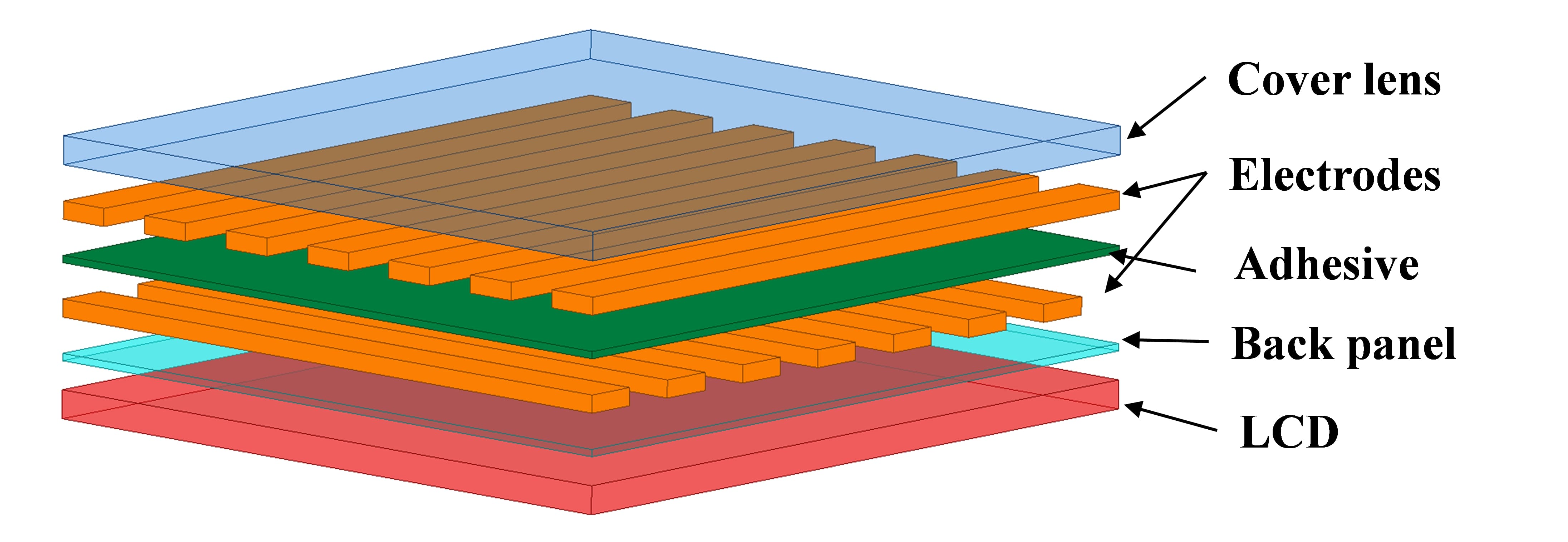}%
\caption{A typical capacitance touchscreen structure. \label{fig:touchscreen_structure}}%
\end{figure}

Among all touchscreen sensing technologies, the capacitive touchscreen is the most popular because it provides a more pleasant user experience and is cost effective. A typical capacitive sensing touchscreen is shown in Fig.~\ref{fig:touchscreen_structure}. There is an array of electrodes under the cover lens of the touchscreen with an adhesive layer between the electrodes that provides mechanical support as well as insulation. The back panel provides insulation between the electrodes and the liquid crystal display (LCD) screen. The electrodes,
adhesive, and back panel are made with optically transparent material. The cover lens is usually made of glass and protects the electrode and the circuit~\cite{wang2011projected}. When the touchscreen is on, a driver circuit delivers a voltage between the two layers of electrodes. The electric field between the two layers of electrodes is constantly sensed. When a person makes contact with the touchscreen, the electric field between the electrode layers are disturbed by their impedance.
Touch events are recognized by sensing this disturbance in the electric field. 

Capacitive sensing touchscreens have already been targeted by several attacks, however, the
majority of touchscreen attacks are passive attacks, e.g., inferring keystrokes~\cite{cai2011touchlogger,aviv2012practicality,owusu2012accessory,xu2012taplogger,miluzzo2012tapprints}, revealing the content on the
touchscreen~\cite{hayashi2014threat,genkin2019synesthesia,li2020wavespy}, etc.
Compared to passive touchscreen attacks, active
attacks~\cite{maruyama2017poster,maruyama2019tap} that manipulate the
touchscreen content and/or events are rare, uncontrolled, and typically require the support of a human touch.

In this paper, we present an active touchscreen attack requiring no physical contact using radiated intentional electromagnetic interference (IEMI). It is the first radiated IEMI touchscreen attack capable of stably recreating complex multi-touch and omni-directonal swipe gestures. Recent work~\cite{wangghosttouch} presents a synchronization-based IEMI touchscreen injection attack and demonstrates several practical attack scenarios. However, because of their reliance on synchronization their range of injected touch events is significantly limited. We also find, see Section~\ref{sec:phone_locator} and Appendix~\ref{app:scanning}, that both the implementation of synchronization and scanning vary by device making the attack difficult to generalize. On the other hand, our attack does not rely on synchronization or the implementation details of scanning to inject stable short-tap, long-press, and omni-directional swipe touch events. This is due in part because we specifically tie the working theory of capacitive touchscreen technology to radiated IEMI electric field strength and signal frequency to precisely and reliably control injected touch events. This in depth analysis allows fully understanding the characteristics of the IEMI disturbance interpreted by the touchscreen as a \textit{human touch}. 

The main contributions of the paper are listed as follows.

\begin{itemize}
    \item We present the underlying mechanism of IEMI based attacks
    on modern capacitive touchscreens. %
    
    \item The principle of IEMI touchscreen attacks is disclosed both theoretically and
    empirically. Crucial factors that influence the effectiveness, including the
    magnitude, frequency, phase, and duration are elaborated.
    
    \item We present an IEMI touchscreen attack capable of injecting both accurate and complex touch events and gestures such as short-tap, long-press, and omni-directional swipes mimicking a \textit{human touch}. \footnote{Readers can find recorded attack videos by
visiting \url{https://invisiblefinger.click/}.}. 
    
    \item We demonstrate practical IEMI touchscreen attacks by designing and implementing an antenna array, screen locator, and injection detector to bridge the gap between simple touch event generation and real-world IEMI attack scenarios. We show and evaluate several practical attacks using multiple commercial devices under different attack scenarios.

\end{itemize}

\section{Background} \label{sec:background}

In this section, we review background knowledge on the sensing strategy of
capacitive touchscreens with a simplified touchscreen model.

\subsection{Capacitive Touchscreens}\label{subsec:capactive_touchscreen}

\begin{figure}[ht]
    \centering
    \includegraphics[width=0.98\columnwidth]{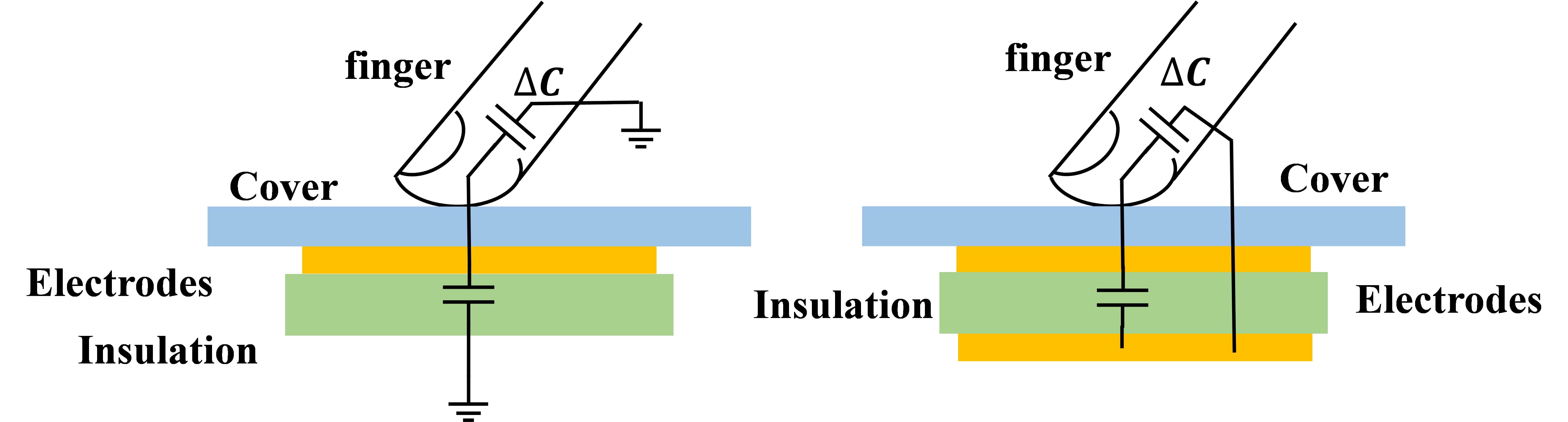}%

    \subfloat[\label{fig:electrode_sensors_a}]{\hspace{.5\linewidth}}
    \subfloat[\label{fig:electrode_sensors_b}]{\hspace{.5\linewidth}}
    \caption{Electrode sensors in capacitance touchscreens: (a) self-capacitance screen; (b) mutual capacitance screen. \label{fig:electrode_sensors}}%
\end{figure}

There are two types of capacitive touchscreens which are widely
used~\cite{barret2010projected}, self-capacitance touchscreens and mutual capacitance touchscreens, shown in
Fig.~\ref{fig:electrode_sensors_a} and Fig.~\ref{fig:electrode_sensors_b} respectively. The $\Delta C$ represents the capacitance
change in the presence of a human finger. When $\Delta C$ is sensed, a touch event is recognized~\cite{luo2012compressive}. 

The self-capacitance touchscreen has a disadvantage because it cannot recognize diagonal touches. In consumer electronics, the ability to sense multi-touch events is beneficial. In contrast, the mutual capacitance touchscreen can
sense several simultaneous touches~\cite{barret2010projected}. Therefore, the mutual capacitance touchscreen is more popular in consumer electronics~\cite{du2016overview}. In this paper, we mainly discuss the mutual capacitance touchscreen although our attack method can also be applied to the self-capacitance touchscreen without loss of generality.

\subsection{Mutual Capacitance Touchscreen}

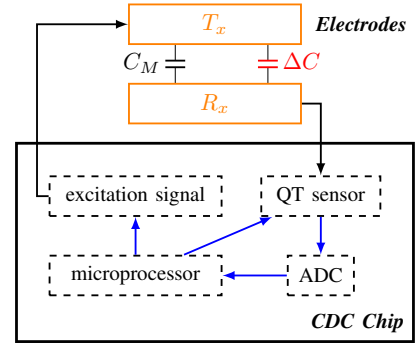
\begin{figure}[ht]
    \centering
    \resizebox{.6\columnwidth}{!}{%
    \begin{tikzpicture}
         \draw (4.4,4.1) to [/tikz/circuitikz/bipoles/length=0.5cm, C, l=$C_M$] (4.4,4.7);
         \draw (5.8,4.7) to [/tikz/circuitikz/bipoles/length=0.5cm, C, color=red, dashed, l=$\Delta C$, red] (5.8,4.1);
    
        \node (up) at (3.8,1.2) [draw,thick,dashed,minimum width=2.6cm, minimum height=0.6cm] {\small microprocessor};
        \node (es) at (3.8,2.4) [draw,thick,dashed,minimum width=2.6cm, minimum height=0.6cm] {\small excitation signal};
        \node (adc) at (6.6,1.2) [draw,thick,dashed,minimum width=1cm, minimum height=0.6cm] {\small ADC};
        \node (qt) at (6.6,2.4) [draw,thick,dashed,minimum width=1.8cm, minimum height=0.6cm] {\small QT sensor};
    
        \draw[>=latex, ->, blue, thick] (up) -- (es);
        \draw[>=latex, ->, blue, thick] (up) -- (qt);
        \draw[>=latex, ->, blue, thick] (qt) -- (adc);
        \draw[>=latex, ->, blue, thick] (adc) -- (up);
    
        \node (tx) at (5,5) [draw,thick,orange,minimum width=2.6cm, minimum height=0.6cm] {$T_x$};
        \node (rx) at (5,3.8) [draw,thick,orange,minimum width=2.6cm, minimum height=0.6cm] {$R_x$};
    
        \node[thick, black, anchor=west] (ele) at (6.4,5){\small \textbf{\textit{Electrodes}}};
    
        \draw[>=latex, ->, black, thick] (es.west) -- +(-5pt,0) |- (tx);
        \draw[>=latex, ->, black, thick] (rx) -| (qt);

        \draw[very thick,black] (2,3.2) rectangle (8,0.2)
            node[above left,black,]{\small \textbf{\textit{CDC Chip}}};
    \end{tikzpicture}%
    }
    \caption{A typical structure of a mutual capacitance touchscreen sensing system. \label{fig:mutual_capacitance_touchscreen}}%
\end{figure}

\begin{figure}[ht]
\centering
\includegraphics[width=0.7\columnwidth]{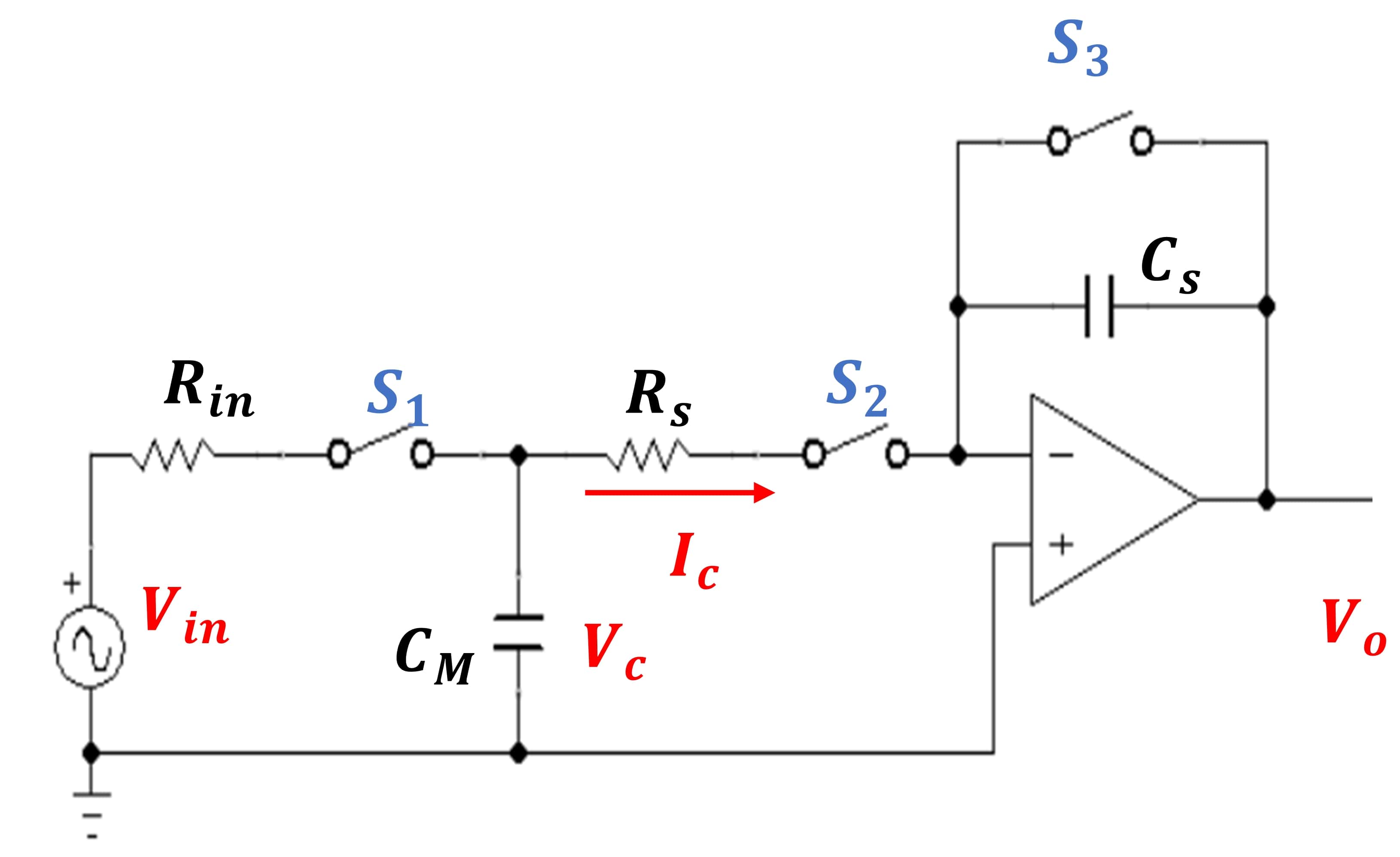}%
\caption{Typical charge transfer circuit topology. \label{fig:charge_transfer_circuit}}%
\end{figure}

A typical structure of a mutual capacitance touch screen system is shown in Fig.~\ref{fig:mutual_capacitance_touchscreen}. The system consists of transmitter (Tx) and
receiver (Rx) electrodes as well as a capacitance to digital converter (CDC)
chip. In the CDC chip, the capacitance between the electrodes is measured with a
charge transfer (QT) sensor. The circuit topology of a QT sensor with an
integrator is shown in Fig.~\ref{fig:charge_transfer_circuit}. The QT sensor
converts the measured capacitance to an analog voltage signal that is then
converted to a digital signal by an analog to digital converter (ADC). A
microprocessor will read in and process the converted digital signal.

\begin{figure}[ht]
    \centering
    \begin{subfigure}[b]{\columnwidth}
        \begin{tikzpicture}%
            \begin{axis}[
                width=\linewidth,height=3cm,
                grid=both,
                xticklabels=\empty,
                enlarge x limits={abs=-1.5},
                legend style={at={(0.0,.91)},anchor=west},
                ]
                \addplot[color=black,very thick] coordinates {
                (1.5, 0)(2, 0)(2, 1)(2.45, 1)(2.45, 0)
                      (3, 0)(3, 1)(3.45, 1)(3.45, 0)
                      (4, 0)(4, 1)(4.45, 1)(4.45, 0)
                      (4.5, 0)
                };
                \node[anchor=west,red] (s0) at (axis cs:1.7,0.5){$S_1$};
                \node[anchor=west] (source) at (axis cs:1.9,0.5){};
                \node (destination) at (axis cs:2.5,0.5){};
                \draw[thick,blue,>=latex,<->](source)--(destination)
                    node[right,black,align=center, inner sep=-1]{\small charging\\\small period};
            \end{axis}
        \end{tikzpicture}    
    \end{subfigure}
    \begin{subfigure}[b]{\columnwidth}
        \begin{tikzpicture}%
            \begin{axis}[
                width=\linewidth,height=3cm,
                grid=both,
                xticklabels=\empty,
                enlarge x limits={abs=-1.5},
                legend style={at={(0.0,.91)},anchor=west},
                ]
                \addplot[color=black,very thick] coordinates {
                (1.50, 0)(1.55, 0)(1.55, 1)(1.9, 1)(1.9, 0)
                         (2.55, 0)(2.55, 1)(2.9, 1)(2.9, 0)
                         (3.55, 0)(3.55, 1)(3.9, 1)((3.9, 0)
                (4.5, 0)
                };
                \node[anchor=west,red] (s2) at (axis cs:2.15,0.5){$S_2$};
                \node[anchor=west] (source) at (axis cs:2.45,0.5){};
                \node (destination) at (axis cs:2.95,0.5){};
                \draw[thick,blue,>=latex,<->](source)--(destination)
                    node[right,black,align=center, inner sep=-1]{\small sensing\\\small period};
            \end{axis}
        \end{tikzpicture}    
    \end{subfigure}
    \begin{subfigure}[b]{\columnwidth}
        \begin{tikzpicture}%
            \begin{axis}[
                width=\linewidth,height=3cm,
                grid=both,
                enlarge x limits={abs=-1.5},
                legend style={at={(0.0,.91)},anchor=west},
                ]
                \addplot[color=black,very thick] coordinates {
                (1.5, 0)(2, 0)(2, 1)(2.05, 1)(2.05, 0)
                      (3, 0)(3, 1)(3.05, 1)(3.05, 0)
                      (4, 0)(4, 1)(4.05, 1)(4.05, 0)
                      (4.5, 0)
                };
                \node[anchor=west,red] (s3) at (axis cs:2.4,0.5){$S_3$};
                \node[anchor=west] (source) at (axis cs:2.65,0.5){};
                \node (destination) at (axis cs:3.05,0.5){};
                \draw[thick,blue,>=latex,->](source)--(destination)
                    node[right,black,align=center, inner sep=-1]{};
                \node[anchor=west] (source1) at (axis cs:2.95,0.5){};
                \node (destination1) at (axis cs:3.35,0.5){};
                \draw[thick,blue,>=latex,<-](source1)--(destination1)
                    node[right,black,align=center, inner sep=-1]{\small reset};
            \end{axis}
        \end{tikzpicture}    
    \end{subfigure}
    \caption{Control signals of the switches $S_1$, $S_2$, and $S_3$. \label{fig:control_signal}}%
\end{figure}
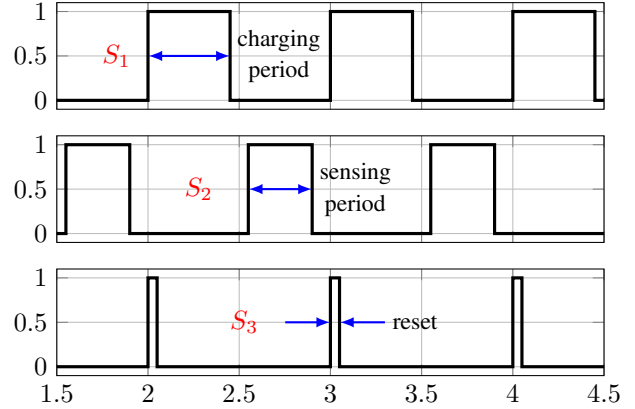

During normal operation, the microprocessor controls three switches, $S_1$,
$S_2$, and $S_3$ (see Fig.~\ref{fig:charge_transfer_circuit}).
Fig.~\ref{fig:control_signal} gives an example of how the control signals are
switched periodically. When the switch $S_1$ is closed, $S_3$ resets $C_s$ and
the excitation signal $V_{in}$ charges the mutual capacitance $C_M$. During this
charging period, the switches $S_2$ and $S_3$ are open and the voltage $V_c$
across $C_M$ is calculated as follows.

\begin{equation}
    V_c=V_{in}\cdot\left(1-e^{-\frac{1}{R_{in}C_M}t}\right) \label{eqn:eqn1}
\end{equation}

\noindent After $C_M$ is charged, $S_1$ is opened and $S_2$ is closed. The charge stored
in $C_M$ will be transferred to $C_s$. Assuming an ideal
op-amp, the current flow through $C_M$ and $C_s$ are equal. The current can
be calculated in (\ref{eqn:eqn2}) or (\ref{eqn:eqn3}).

\begin{equation}
    I_c=-C_M\frac{dV_c}{dt} \label{eqn:eqn2}
\end{equation}

\begin{equation}
    I_c=-C_s\frac{dV_o}{dt} \label{eqn:eqn3}
\end{equation}

\noindent
By solving and integrating (\ref{eqn:eqn2}) and (\ref{eqn:eqn3}) simultaneously
over the time with initial conditions, the output voltage $V_o$ is derived in
(\ref{eqn:eqn4}).

\begin{equation}
    V_o=-\frac{C_M}{C_s}V_c \label{eqn:eqn4}
\end{equation}

\noindent
Based on (\ref{eqn:eqn4}), the mutual capacitance $C_M$ can be calculated from
$V_o$. When the sensing period is completed, at the beginning of the next
period, $C_s$ is discharged by closing $S_3$.

When a touch event occurs, $C_M$ is changed by $\Delta C$ due to the presence
of a human finger. This change can be either positive or
negative~\cite{hwang2010highly} depending on human impedance
variations~\cite{hayashi2013efficient}. The output voltage can be calculated as follows when the touch event occurs.

\begin{equation}
    V_{oT}=-\frac{(C_M\pm \Delta C)}{C_s} V_c=V_o+V_T \label{eqn:eqn5}
\end{equation}

\noindent where $V_T$ is the output voltage variation and is calculated as follows.

\begin{equation}
    V_T = \pm \frac{\Delta C}{C_s} V_c \label{eqn:eqn6}
\end{equation}

\noindent
A touch event is recognized if the following criterion is met.

\begin{equation}
    |V_T| \geq V_{th} \label{eqn:eqn7}
\end{equation}

\noindent where $V_{th}$ is the threshold voltage. 

The sensing strategy in Fig.~\ref{fig:control_signal} senses and compares the
output voltage to every cycle's threshold voltage. In many applications, a
multi-cycle sensing strategy is usually used to get a more accurate result for
each touch event by measuring $V_o$ and $V_T$ multiple times. In a multi-cycle sensing
strategy, $C_s$ is reset every $N$ cycles. In this way, $V_o$ and $V_T$ are
the sum of the voltages in $N$ cycles. The touch recognition criterion in
(\ref{eqn:eqn7}) in this case is as follows.

\begin{equation}
    |\sum V_T|\geq V_{thN} \label{eqn:eqn8}
\end{equation}

\noindent where $V_{thN}$ is the threshold voltage defined for the $N$ cycle
sensing strategy. If the voltage variations in these cycles are the same, then
we have $\sum V_T=N\cdot V_T$.

Based on (\ref{eqn:eqn1}) - (\ref{eqn:eqn8}), the $\Delta C$ between every pair
of electrodes can be measured by QT sensors. The locations of the electrodes
represent the touchable locations on the touchscreen. 

\section{Threat Model} \label{sec:threatmodel}
In this paper, we assume that the attacker is equipped with tools that can generate IEMI signals including electrode plates, a signal generator and an RF power
amplifier. The electrode plates are used to radiate IEMI signals and can be hidden under a table or desk (check our experimental setup in Section~\ref{sec:evaluation} for more details). We further assume that the victim's device is equipped with a capacitive touchscreen. We do not require the victim to have a certain brand of touchscreen device, nor do we have any limitations on the operating system. We aim to mimic a real
world setting in which a victim puts their smart device on the table under which the electrode plates are attached. We assume the victim puts the smart device face down on the table, a typical way to prevent screen
eavesdropping. The attack does not need to have prior knowledge of the phone location or orientation. The attacker can use the electrode plates to generate a precise touch event on the screen and further manipulate the victim device to perform security oriented attacks, such as connecting to Apple headphones to remotely control the victim device, or installing malicious applications.

\section{IEMI Attack Preliminaries} \label{sec:attackmethod}

In this section, we will present the fundamental electromagnetic concepts and derive the corresponding circuit model of the touchscreen under the IEMI attack. The concept and the model here pave the way to systematically analyze the behavior of a touchscreen under IEMI attacks. 

\subsection{IEMI Attack Intuition}
From Section \ref{sec:background}, we learned that a touch event is sensed if the output
voltage variation, $V_T$, is larger than the threshold voltage, $V_{th}$. Therefore, a ghost touch event can be induced when a radiated IEMI signal causes $V_o$ to exceed the threshold voltage, which allows attackers to control the device without physically touching the screen.

\subsection{Generating a Targeted Radiated IEMI Signal}

There are multiple ways to generate the radiated IEMI signal. A simple and
straightforward method is to generate an electric field using two electrode
plates that are facing each other. It is also possible to generate the
electric field with phased antenna arrays where the direction of the IEMI is
controlled by the array factor. The third method is to leverage directional
antennas, such as Log-periodic antennas or Yagi-Uda~\cite{oldenburg_2019}
antennas.

Based on our attacking principle analysis later in this paper, electrodes
(near-field antenna) are more suitable for existing smart touchscreen enabled
electronic devices, therefore, our work focuses on an electrode-based IEMI attack
and we will show that only one electrode is enough to perform an attack. For
convenience, we simply call an electrode (a near-field antenna) as an antenna in
later analysis.

\subsection{Effect of Radiated IEMI on a Touchscreen}
\begin{figure}[ht]
\centering
\includegraphics[width=0.98\columnwidth]{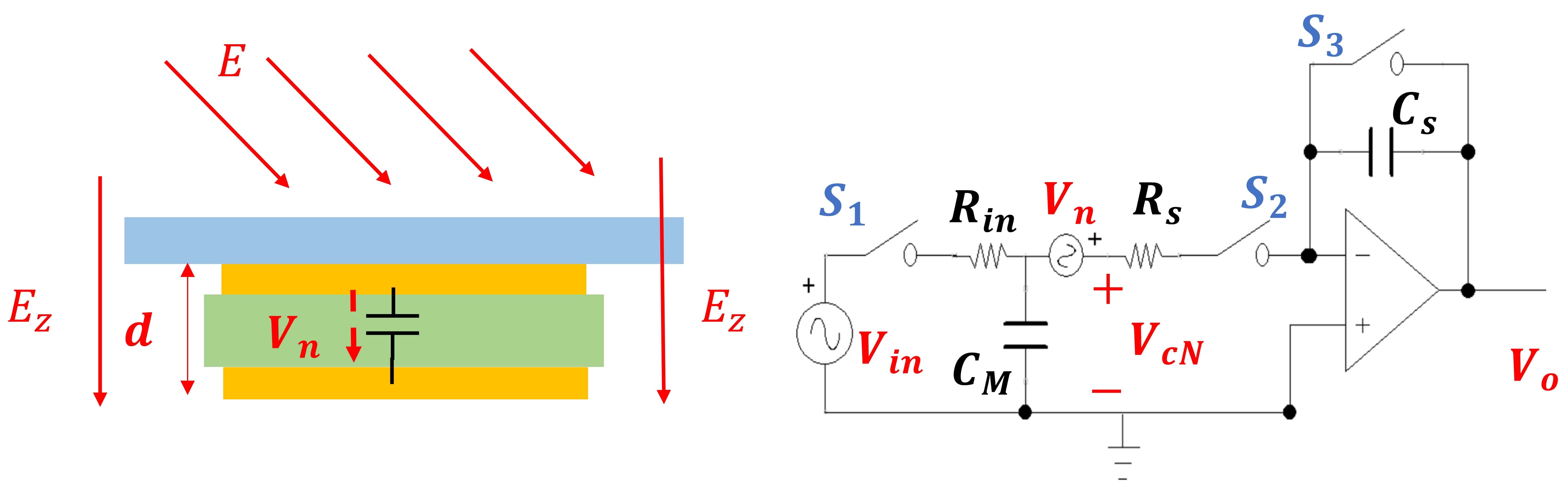}%
\vspace{-1em}
\subfloat[\label{figs:E_field_interference_a}]{\hspace{.5\linewidth}}
\subfloat[\label{figs:E_field_interference_b}]{\hspace{.5\linewidth}}
\caption{Illustration of the E field interference: (a) E field on touchscreen electrodes and (b) equivalent circuit of QT Sensor. \label{fig:E_field_interference}}%
\end{figure}

Fig.~\ref{fig:E_field_interference} depicts the electric field (referred to as E field hereafter) interference due to an external E field on a touchscreen, and its effect on the equivalent QT sensor circuit. The presence of an external E field induces a displacement current that flows through and adds or removes charge from the mutual capacitance touchscreen electrodes. Note that $V_o$ of the QT sensor depends on the total charge stored in the mutual
capacitance $C_M$. Thus, the measured output voltage variation $V_T$ is controlled by the targeted E field and can induce ghost touches.

\subsection{Relationship of IEMI E Field Strength and Touchscreen Attack }
\label{subsec:efield_strength}

To introduce a touch event with an IEMI attack, the E field strength needs to meet certain requirements. The E field interference on a touchscreen is shown in Fig.~\ref{figs:E_field_interference_a}. The critical E field that is required to cause a ghost touch is defined as $E_{crit}$ and can be calculated as
follows. The detailed derivation process can be found in Appendix~\ref{app:efield}.

We assume $V_{Tn}$ is the output voltage variation caused by the IEMI noise. To generate the ghost touch, we need to fulfill the following requirement, i.e.,

\begin{equation}
    \left|V_{T n}\right| \geq\left|V_{T}\right|=\frac{\Delta C}{C_{s}} V_{c}=\frac{Q_{t}}{C_{s}}   \label{eqn:eqn15}    
\end{equation}

\noindent where $Q_t=\Delta C\cdot V_c$, representing the charge change caused
by the real touch. Solving (\ref{eqn:eqn12}), (\ref{eqn:eqn14}) and
(\ref{eqn:eqn15}) simultaneously, 

\begin{equation}
    E_{crit}=\frac{Q_t}{\varepsilon_0\cdot\varepsilon_r\cdot A} \label{eqn:eqn16}
\end{equation}

\noindent
Based on (\ref{eqn:eqn16}), if $E_Z$ is larger than $E_{crit}$, a ghost touch is
successfully generated.

\begin{figure}[ht]
\centering
\includegraphics[width=0.98\columnwidth]{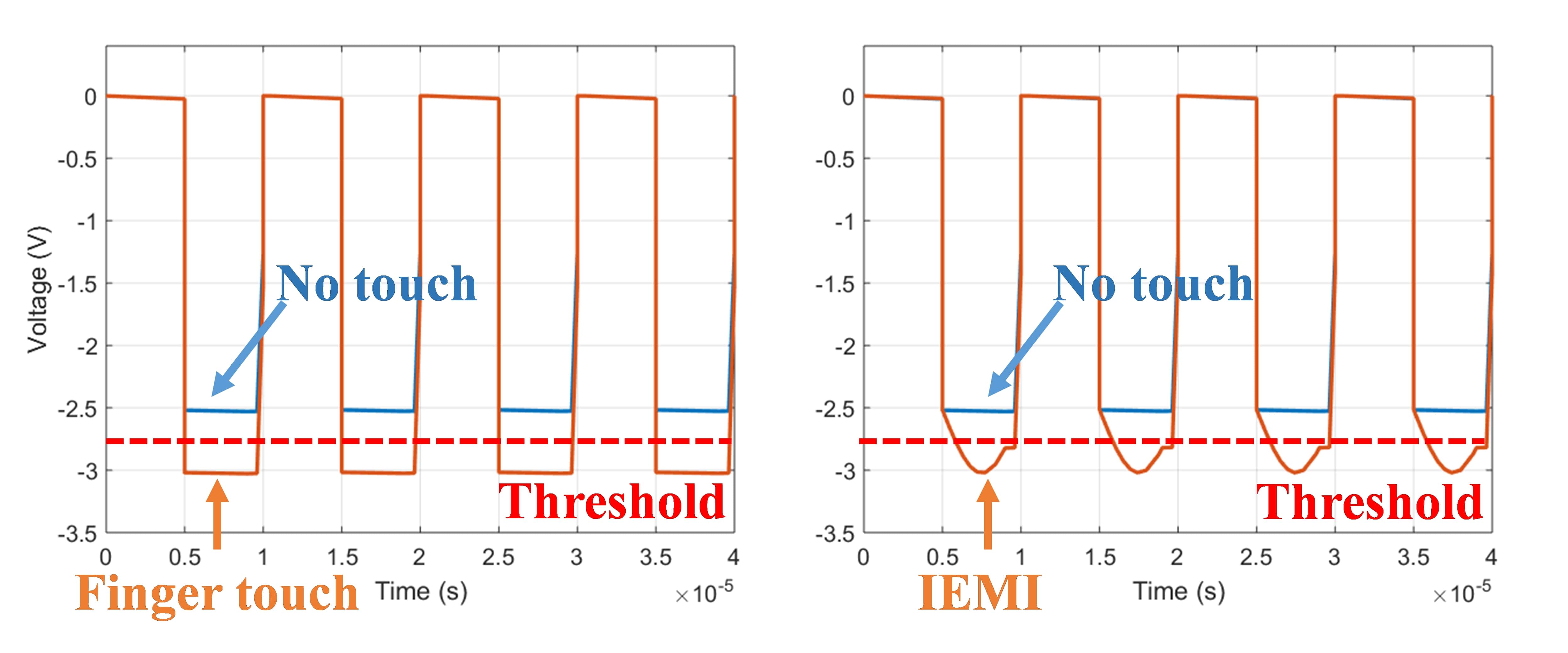}%
\vspace{-1em}
\subfloat[\label{figs:simulated_output_voltage_a}]{\hspace{.5\linewidth}}
\subfloat[\label{figs:simulated_output_voltage_b}]{\hspace{.5\linewidth}}
\caption{Simulated output voltage of a QT sensor: (a) output voltage with a
finger touch and (b) output voltage under IEMI with the critical E field
strength.  \label{fig:simulated_output_voltage}}
\end{figure}

\medskip
\noindent\textbf{Simulation Validation of Touchscreen Response to Radiated IEMI: }Fig.~\ref{figs:simulated_output_voltage_a} and~\ref{figs:simulated_output_voltage_b}
show the simulated $V_o$ of a single QT sensor under a finger touch and IEMI attack based on the developed model, respectively. For this simulation, switches $S1$-$S3$ are
controlled with 100kHz signals as shown in Fig.~\ref{fig:control_signal}. All
simulation parameters are listed in Table~\ref{tab:tab1}. The touch event is
simulated using a positive 0.5 pF capacitance change. The IEMI signal is simulated using
a noise voltage source $V_n$ at the input of the QT sensor. $V_{th}$ is set to 2.75 V. To cause a ghost touch, $V_n$ should meet the requirement in \ref{eqn:eqn17}.

\begin{equation}
    V_{n} \geq V_{in} \cdot \frac{\Delta C}{C_{M}} \label{eqn:eqn17}
\end{equation}

As shown in
Fig.~\ref{figs:simulated_output_voltage_a}, $V_o$ changes when there is a finger touch due to the change in capacitance. Once $V_o$ exceeds
$V_{th}$, a touch event is recognized. Under the simulated IEMI attack (shown in Fig.~\ref{figs:simulated_output_voltage_b}), $V_o$ exceeds $V_{th}$ even when there is no touch. This validates our QT sensor model analysis, and motivates our subsequent experiments for generating ghost touch events in real scenarios.

\begin{table}[htbp]
\centering
\caption{QT Sensor Simulation Parameters} \label{tab:tab1}
\begin{tabular}{||c|c|c|c||}
\hline\hline
Parameter & Value & Parameter & Value  \\\hline
$V_{in}$ & 5 V & $C_M$ & 3 pF  \\\hline
$R_{in}$ & 1 $\Omega$ & $C_s$ & 10 pF  \\\hline
$R_s$ & 1 $\Omega$ & $\Delta C$ & 0.5 pF  \\\hline
$V_{th}$ & 2.75 V & $V_n$ & 0.8V/100kHz \\\hline\hline
\end{tabular}
\end{table}

\subsection{Relationship of IEMI Frequencies and a Successful Attack} \label{subsec:freq_analysis}

\begin{figure}[ht]
\centering
\includegraphics[width=0.98\columnwidth]{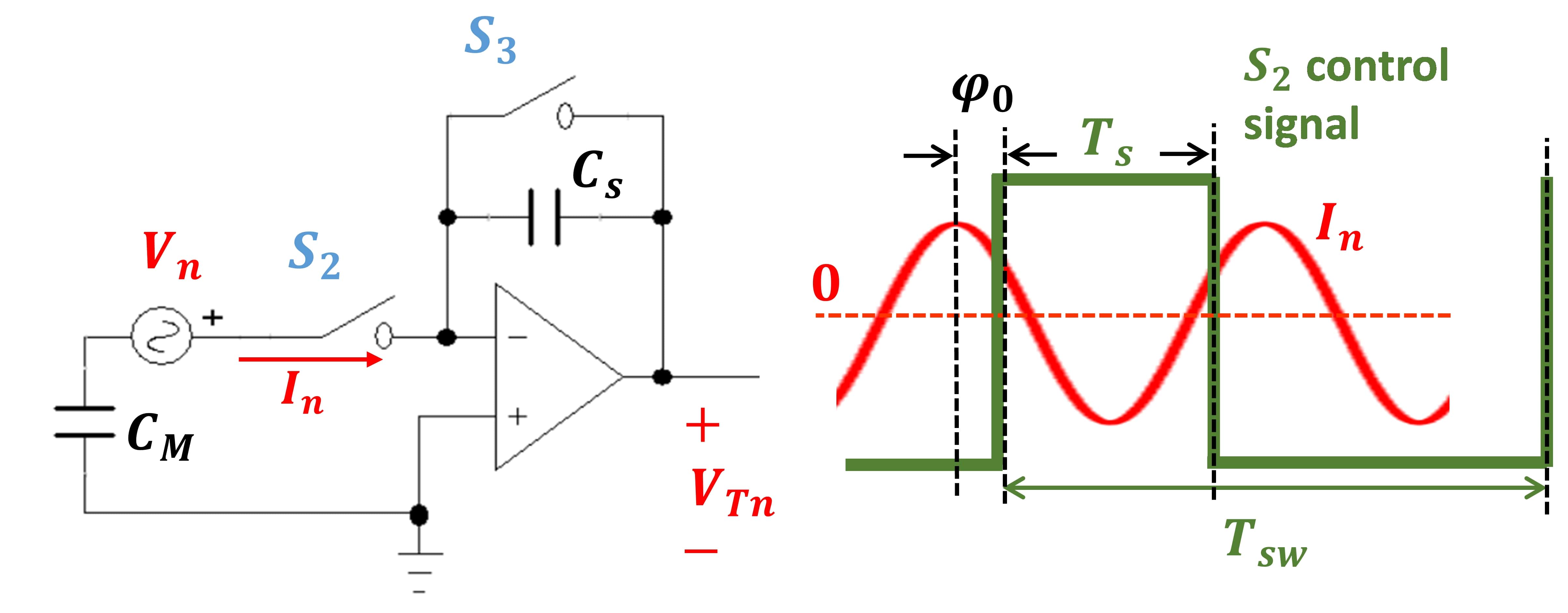}%
\vspace{-1em}
\subfloat[\label{figs:equivalent_circuit_a}]{\hspace{.5\linewidth}}
\subfloat[\label{figs:equivalent_circuit_b}]{\hspace{.5\linewidth}}
\caption{(a) Equivalent circuit of a QT sensor in a touchsreen controller and (b) $S_2$ control signal and $I_n$ waveforms. \label{fig:equivalent_circuit}}%
\end{figure}

From Section~\ref{subsec:efield_strength}, we know that the E field strength will, in part,
decide the IEMI attack effectiveness. Nevertheless, as shown in previous work
\cite{savage2012overview}, the frequency of the interfering signal also plays a
critical role. Therefore, we conduct the following analysis to first reveal the
relationship of IEMI frequencies and a successful IEMI attack.
Fig.~\ref{figs:E_field_interference_b} shows the voltage source $V_n$ which is
the input voltage of the QT sensor due to the IEMI attack. Based on the
superposition theory, we can derive the equivalent circuit of a QT sensor under an
IEMI attack where only the noise source $V_n$ is considered (see
Fig.~\ref{figs:equivalent_circuit_a}). $R_s$ is ignored since it is much smaller
than the impedance of $C_M$.

\noindent
The mathematical calculation of the minimum IEMI interference that can
cause a ghost touch event is thoroughly explained in Appendix~\ref{app:iemi_fre}. The calculation gives us the lower boundary of IEMI attacks. In real
attacks, we would like to maximize the IEMI interference. A similar calculation
process also applies. The maximum interference can be achieved if one
of the following two conditions is met.

\medskip
\begin{itemize}[leftmargin=*]
\item \textit{Condition 1:} The phase angle is $\varphi_0=\frac{3\pi}{2}$ and the
frequency of the IEMI signal satisfies (\ref{eqn:eqn27}) and (\ref{eqn:eqn28})
simultaneously. 

\begin{equation}
    f_E=\frac{f_{sw}}{4D_s}+\frac{kf_{sw}}{D_s} \;\;\;\;\;      k=0,1,2,3,\dots
    \label{eqn:eqn28}
\end{equation}

\item \textit{Condition 2:} The phase angle is $\varphi_0=\frac{\pi}{2}$ and the
frequency of the IEMI signal satisfies (\ref{eqn:eqn27}) and (\ref{eqn:eqn29})
simultaneously. 
\end{itemize}

\begin{equation}
    f_E=\frac{{3f}_{sw}}{4D_s}+\frac{kf_{sw}}{D_s} \;\;\;\;\;      k=0,1,2,3,\dots
    \label{eqn:eqn29}
\end{equation}

\noindent
As we will show in Section \ref{sec:poc_quantitative}, by conducting several
experiments with a Chromebook equipped with a touchscreen diagnostic data collection
program, we confirm our developed theory by identifying various frequencies at which
ghost touches are caused at the required minimum E field. The impact of $\varphi_0$ is minimized by finding the worst case in multiple measurements at
each frequency.

\section{Proof-of-Concept Evaluation} \label{sec:poc}

In Section~\ref{sec:attackmethod}, we developed a theory for IEMI ghost touch attacks and validated it using simulations. In this section, we will demonstrate the IEMI attack using a relatively ideal experiment setup by targeting a laptop with electrode plates placed directly on both sides of the laptop touchscreen.
With this setup, we generate real experimental results to validate our previous analysis, e.g., the required E field and needed frequencies for effective IEMI attack signals.

\subsection{Experimental Setup}

As a proof-of-concept, we generate radiated IEMI using electrode plates
placed on opposite sides of our target device. A signal generator (RIGOL DS 1052E) and an
RF power amplifier (Amplifier Research 25A250A) are used to generate the
desired voltage. The output of the RF amplifier is monitored by an
oscilloscope (RIGOL MSO4054). The touchscreen of a Chromebook laptop is
used as the target. This laptop is installed with Touch Firmware
Tests~\cite{touchfirmwaretests} developed by the Chromium Project. This
program records all of the touched positions recognized by the touchscreen
controller during the test. The recorded data is collected by an external
device over Wi-Fi. A test report is also generated that lists all touched
locations during the testing period. During the test, the Chromebook is
disconnected from the adapter and placed on a non-conductive surface 70 cm
above the ground to avoid undesired EMI noise.

\subsection{IEMI Generation}

\begin{figure}[ht]
\centering
\includegraphics[width=0.98\linewidth]{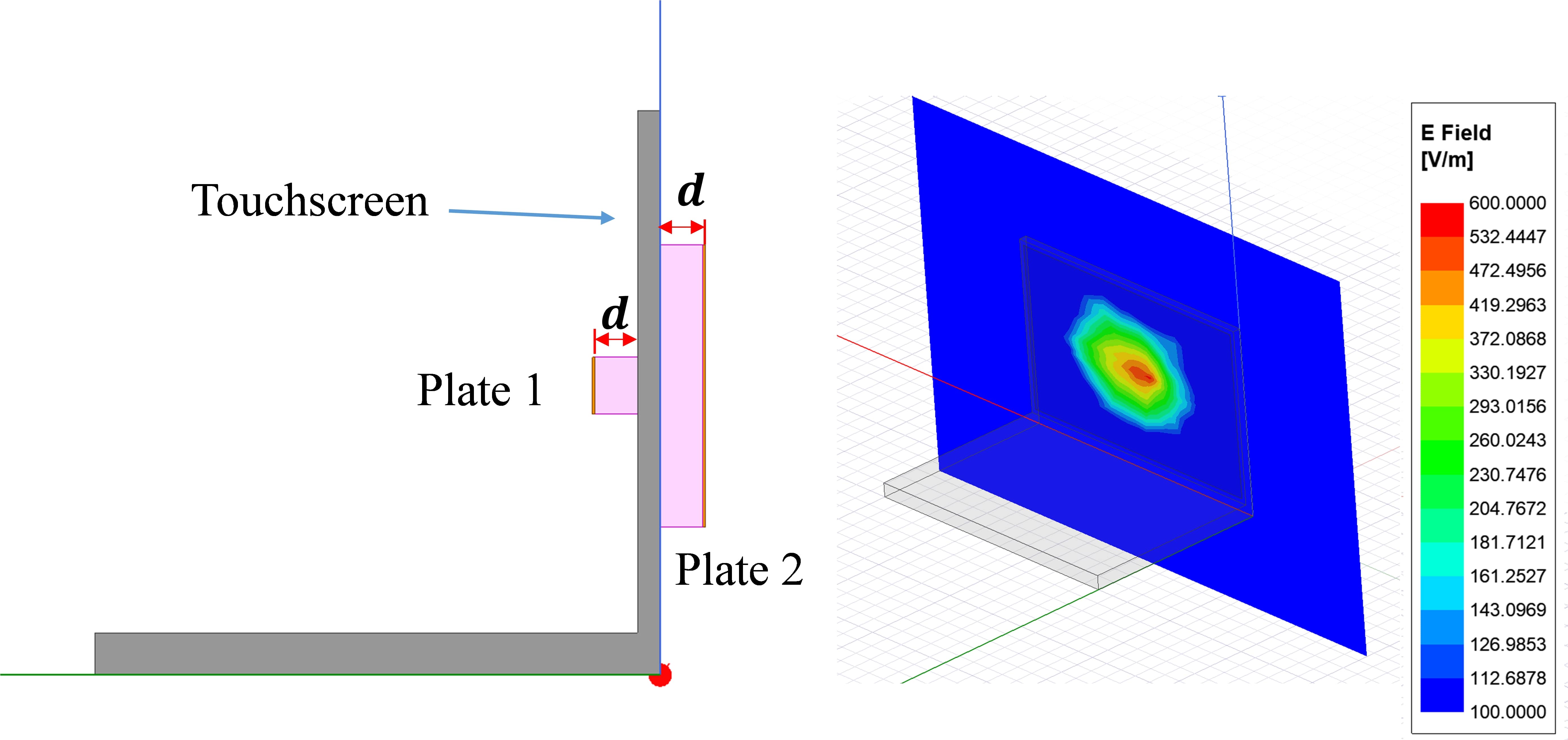}%
\vspace{-1em}
\subfloat[\label{figs:electric_field_simulation_a}]{\hspace{.5\linewidth}}
\subfloat[\label{figs:electric_field_simulation_b}]{\hspace{.5\linewidth}}
\caption{Electric field simulation: (a) cross-sectional view and (b) simulated electric field on the surface of the touchscreen. \label{fig:electric_field_simulation}}%
\vspace{-.5em}
\end{figure}

The E field parameters are selected based on our calculations in
Section~\ref{subsec:freq_analysis}. Fig.~\ref{fig:electric_field_simulation} shows the placement of the two electrode plates. Plate 1 is an 8 mm x 8 mm copper plate taped on the front of the touchscreen. Plate 2 is a 150 mm x 150 mm copper plate taped on the back of the touchscreen. The distances $d$ between each plate and the touchscreen are both 10 mm (see Fig.~\ref{figs:electric_field_simulation_a}). A
non-conductive foam sheet is inserted between the plates and the touchscreen for mechanical support. The thickness $t$ of the touchscreen itself is 5 mm.  
The dielectric constant of the foam sheet is in the range of 1.8 -
3~\cite{mohamed2016frequency}. To simplify the calculation of E field strength,
$E_z$, we use the following
equation based on $V_E$, the voltage across the plates.

\begin{equation}
    E_z=\frac{V_E}{2d+t} \label{eqn:eqn18}
\end{equation}

\noindent
Further, to validate the accuracy of (\ref{eqn:eqn18}), we compare our calculated results with simulation results using Ansys HFSS~\cite{hfss_2021}. Note that the simulation
reflects the real configuration by considering the foam sheet and the plate sizes. The HFSS uses finite element analysis to solve Maxwell's
equation, thereby providing accurate calculation results.

Fig.~\ref{figs:electric_field_simulation_b} shows the simulated E field on the touchscreen caused by the two plates when $V_E=15 V$. We found that the magnitude of the simulated E field is approximately equal to the calculated results using (\ref{eqn:eqn18}), which indicates that the simplified (\ref{eqn:eqn18}) is a good estimate for the generated E
field strength. Hereafter, we will rely on (\ref{eqn:eqn18}) to derive the
$V_E$ based on the required $E_z$.

\subsection{Evaluation of E Field Strength IEMI on Touchscreen Behavior to Validate Our Theory}

To exclude possible interference from the electrode plates affecting the touchscreen functionality, we first do not apply voltage to the electrode plates and collect touchscreen diagnostic data by drawing a random pattern on the touchscreen with a finger. This confirms that the touchscreen functions normally.

\smallskip
\noindent\textbf{Stationary IEMI attack: }Once we confirm the electrodes themselves have no impact on the touchscreen, we calculate the required $V_E$ for an IEMI attack. We collect parameters for a typical touchscreen from~\cite{barret2010projected}. The minimum detectable capacitance change $\Delta C$ is 0.1 pF and the touchscreen
controller excitation signal $V_{in}$ is 5 V. We also incorporate the overlap area $8 mm \times 8 mm$ due to the electrode. From (\ref{eqn:eqn16}), we have $E_{crit} = 883 V/m$. Following
(\ref{eqn:eqn18}), the corresponding $V_E$ is calculated as 22 V.

\begin{figure}[ht]
\centering
\begin{subfigure}[b]{.47\linewidth}
  \centering
  \begin{tikzpicture}
    \node [
        above right,
        inner sep=0] (image) at (0,0) {\frame{\includegraphics[width=\linewidth]{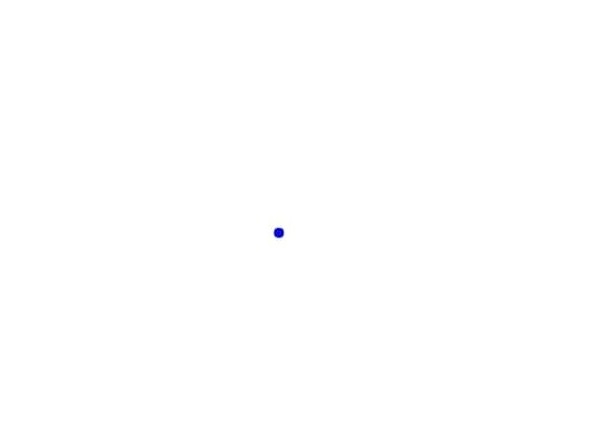}}};
    \begin{scope}[
      x={($0.1*(image.south east)$)},
      y={($0.1*(image.north west)$)}]
      \draw[very thick,orange,dashed] (4,4) rectangle (5.5,6) 
        node[anchor=north,above,black,fill=white, inner sep=5]{\small plate 1 location};
    \end{scope}
  \end{tikzpicture}  
  \caption{}
  \label{fig:fig13a}
\end{subfigure}
\begin{subfigure}[b]{.47\linewidth}
  \centering
  \begin{tikzpicture}
    \node [
        above right,
        inner sep=0] (image) at (0,0) {\frame{\includegraphics[width=\linewidth]{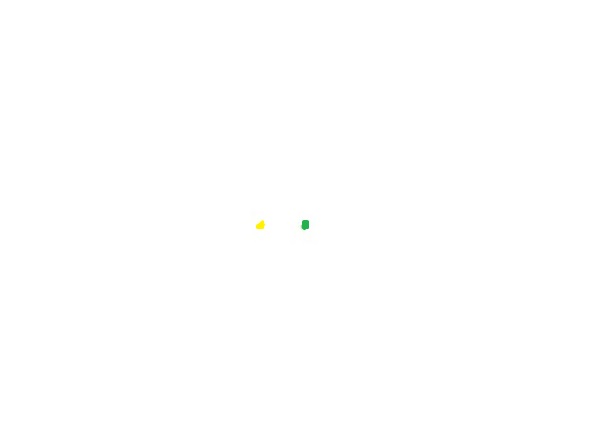}}};
    \begin{scope}[
      x={($0.1*(image.south east)$)},
      y={($0.1*(image.north west)$)}]
      \draw[very thick,orange,dashed] (4,4) rectangle (5.5,6) 
        node[anchor=north,above,black,fill=white, inner sep=5]{\small plate 1 location};
    \end{scope}
  \end{tikzpicture}    \caption{}
  \label{fig:fig13b}
\end{subfigure}
\caption{Ghost touch under an IEMI attack with (a) 20 V, 140 kHz and (b) 25 V 140 kHz voltage excitation $V_E$.}
\label{fig:fig13}
\end{figure}

We then set $V_E$ on the signal generator to be a sinusoidal voltage source with a frequency of 140kHz.
Instead of applying 22 V directly, the amplitude of $V_E$ is gradually increased until a ghost touch is observed. The process is repeated three times to find the minimum voltage that causes the ghost touch. In our experiment, we do not detect ghost touches when $V_E$ is lower than 20 V. When the voltage is higher than 20 V, however, ghost touches start to appear. As shown in Fig.~\ref{fig:fig13a}, a
ghost touch is successfully generated at the center of plate 1 when $V_E$ is 20 V. Note that the required minimum $V_E$ for ghost touches is close to our theoretical calculation (i.e., 22 V), showing that our analysis is accurate. When we increase $V_E$ above 20 V, multiple ghost touches are observed. This is because when the voltage is high compared to the minimum $V_E$, several locations under plate 1 (as opposed to just one) have sufficiently high E field strengths to induce ghost touches.
Fig.~\ref{fig:fig13b} shows that two ghost touches are generated when $V_E$ is 25 V. %

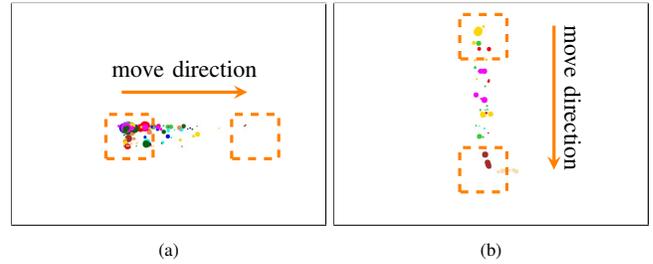
\begin{figure}
\centering
\begin{subfigure}{.47\linewidth}
  \centering
  \begin{tikzpicture}
    \node [
        above right,
        inner sep=0] (image) at (0,0) {\frame{\includegraphics[width=\linewidth]{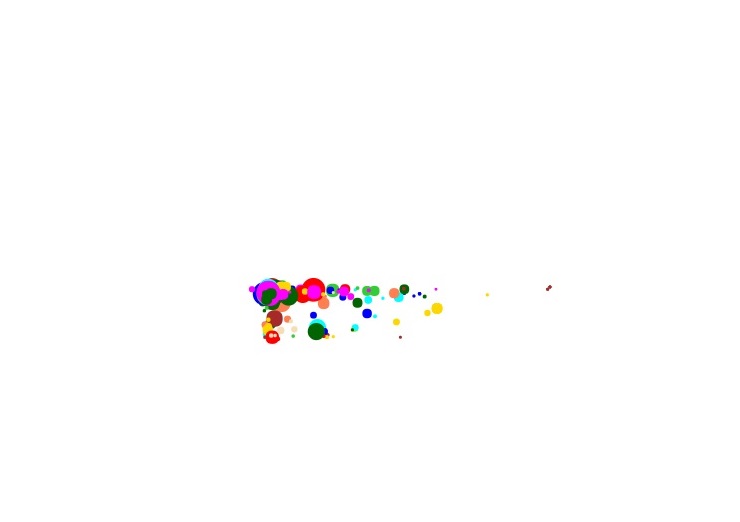}}};
    \begin{scope}[
      x={($0.1*(image.south east)$)},
      y={($0.1*(image.north west)$)}]
       \draw[stealth-, very thick,orange] (7.5,6) -- ++(-4,0) 
       node[above,black,midway, align=center,inner sep=5]{\small move direction};
      \draw[very thick,orange,dashed] (3,3) rectangle (4.5,5);
      \draw[very thick,orange,dashed] (7,3) rectangle (8.5,5);
    \end{scope}
  \end{tikzpicture}
  \caption{}
  \label{fig:fig14a}
\end{subfigure}
\begin{subfigure}{.47\linewidth}
  \centering
  \begin{tikzpicture}
    \node [
        above right,
        inner sep=0] (image) at (0,0) {\frame{\includegraphics[width=\linewidth]{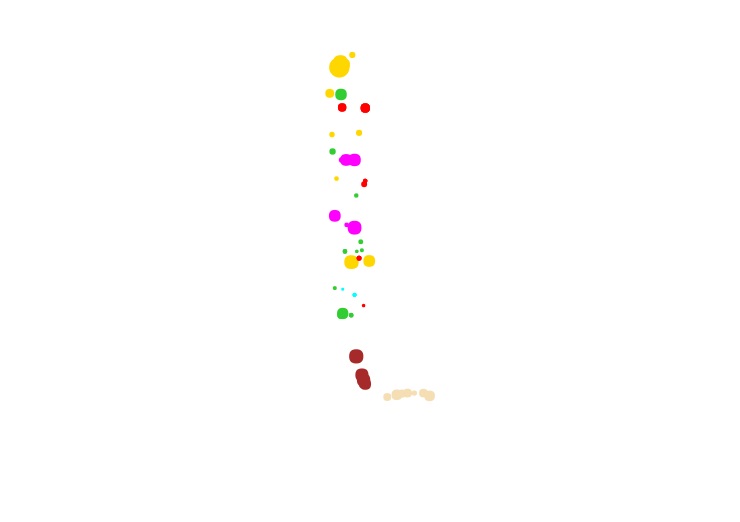}}};
    \begin{scope}[
      x={($0.1*(image.south east)$)},
      y={($0.1*(image.north west)$)}]
       \draw[stealth-, very thick,orange] (7,2.5) -- ++(0,6.5) 
       node[above,black,midway, align=center, rotate=-90]{\small move direction};
      \draw[very thick,orange,dashed] (4,7.5) rectangle (5.5,9.5);
      \draw[very thick,orange,dashed] (4,1.5) rectangle (5.5,3.5);
    \end{scope}
  \end{tikzpicture}
  \caption{}
  \label{fig:fig14b}
\end{subfigure}
\caption{Ghost touchpoints with plate 1 moves (a) from left to right and (b) from top to bottom.}
\label{fig:fig14}
\end{figure}

\smallskip
\noindent\textbf{Moving IEMI attack: }We have demonstrated that the touchscreen is vulnerable to stationary IEMI
sources. We further expand our experiment by moving our electrode plates around
to verify if only certain locations on the touchscreen are vulnerable. To account for jitter caused by moving the electrode plates, we increase the applied $V_E$ to 30V / 140kHz (E field strength of $1200 V/m$) to ensure the E field is always higher than $E_{crit}$. 
As shown in Fig.~\ref{fig:fig14a}, many ghost touch points are evident
when plate 1 moves from left to right. Fig.~\ref{fig:fig14b} shows the ghost touch points when plate 1 moves from top to bottom. The results show that all physical locations of the touchscreen are equally vulnerable to an IEMI attack.

\subsection{Evaluation of IEMI Frequencies on Touchscreen Behavior to Validate Our Theory} \label{sec:poc_quantitative}

As we mentioned in Section~\ref{subsec:freq_analysis}, the E field frequency also impacts the IEMI attack in addition to its strength. We therefore conduct several experiments to validate our analysis on calculating the required signal frequencies for a successful IEMI attack.

\begin{figure}[ht]
\centering
\begin{tikzpicture}%
    \begin{semilogxaxis}[
        xlabel=Frequency (\textit{Hz}),
        ylabel=External E field (\textit{V/m}),
        width=0.98\linewidth,height=6cm,
        ytick={500, 1000, 1500, 2000, 2500, 3000},
        grid=both,
        legend style={at={(0.0,.91)},anchor=west},
        ]

    \addplot[color=red,thick] table {data/e_field.dat};

    \end{semilogxaxis}
\end{tikzpicture}
\caption{Minimum E field that causes the ghost touch at different frequencies \label{fig:minimum_E_field}}%
\end{figure}
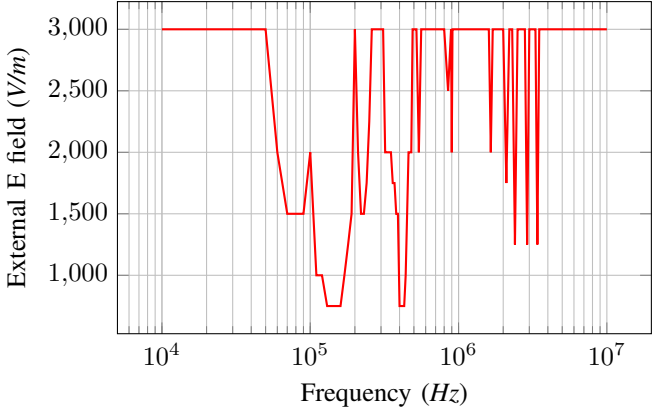

\smallskip
\noindent\textbf{Sweeping IEMI Attack Frequencies to Validate Our Theory: }From~\cite{hayashi2013efficient,zhang2019investigating}, we know that the touchscreen system is sensitive to noise in the range of $100\, kHz$ to $1\, MHz$ due to integrated low pass filters in the touch sensing circuit.
We sweep the frequency from $10\, kHz$ to $10\, MHz$
to cover the sensitive frequency range using steps of $10\, kHz$. With each chosen frequency, we tune the voltage applied on the two electrode plates until ghost touches are detected. If the generated E field exceeds $3000 V/m$ and there is still no ghost touches detected, then we claim that the selected E field frequency cannot generate a ghost touch. We run each test for 5 seconds and after each measurement reboot the Chromebook to reset
the touchscreen. The procedure is repeated three times for each frequency. All collected results are plotted in Fig.~\ref{fig:minimum_E_field} which shows a complete view of the frequency dependency for successful IEMI attacks. As we can see in this figure, certain excitation frequencies out-perform other frequencies (requires smaller E field strength to trigger ghost touch), which validated our previous theory of IEMI frequencies, see equation~(\ref{eqn:eqn28}) and~(\ref{eqn:eqn29}).

\smallskip
\noindent\textbf{Targeted IEMI Attack Frequencies to Validate Our Theory: }In Section \ref{subsec:freq_analysis}, we show that $f_{sw}$ and $D_s$ determine the minimum/maximum IEMI interference using an E field with frequency $f_E$. These parameters can be calculated from two adjacent frequencies with the maximum interference (local lowest $E_{crit}$). Using the results presented in Fig.~\ref{fig:minimum_E_field}, we select two adjacent
frequency points and derive $f_{sw}=70 kHz$ and $D_s=0.125$. Based on these calculations, we can then derive all E field frequencies that
can cause  minimum IEMI interference (denoted as $f_{Emin}$) or maximum IEMI interference (denoted as $f_{Emax}$) using (\ref{eqn:eqn24}), (\ref{eqn:eqn28}) and (\ref{eqn:eqn29}). In the frequency range of $100\, kHz$ to $1\, MHz$, $f_{Emax}$ and $f_{Emin}$ are listed as follows.

$$f_{Emax}=140\, kHz,\, 420\, kHz,\, 700\, kHz,\, 980\, kHz   $$
$$f_{Emin}=\, 560\, kHz,\, 1120\, kHz $$

Note that these calculated frequencies match the experimental results shown in Fig.~\ref{fig:minimum_E_field}. For frequencies other than $f_{Emin}$ and $f_{Emax}$, we can still obverse ghost touches with larger than minimum E field
strengths. It is worth noting that the IEMI signal cannot cause any interference at $700\, kHz$. This is likely caused by internal filters that are in place to avoid undesired interference from internal electronics components at those frequencies. For frequencies higher than 1 MHz, the impact of the sensor circuit's internal low pass filter and parasitic parameters become more significant~\cite{zhang2019investigating}. Since this is often proprietary
information of touchscreen manufacturers, the experimental results become less consistent with our calculations. When we set the frequency larger than 3.4 MHz, no ghost touches are detected.

\section{Precise Screen Control using IEMI Attack} \label{sec:analysis_precise}

\begin{figure}[ht]
    \centering
    \begin{subfigure}[t]{.35\linewidth}
        \centering
        \includegraphics[width=\linewidth]{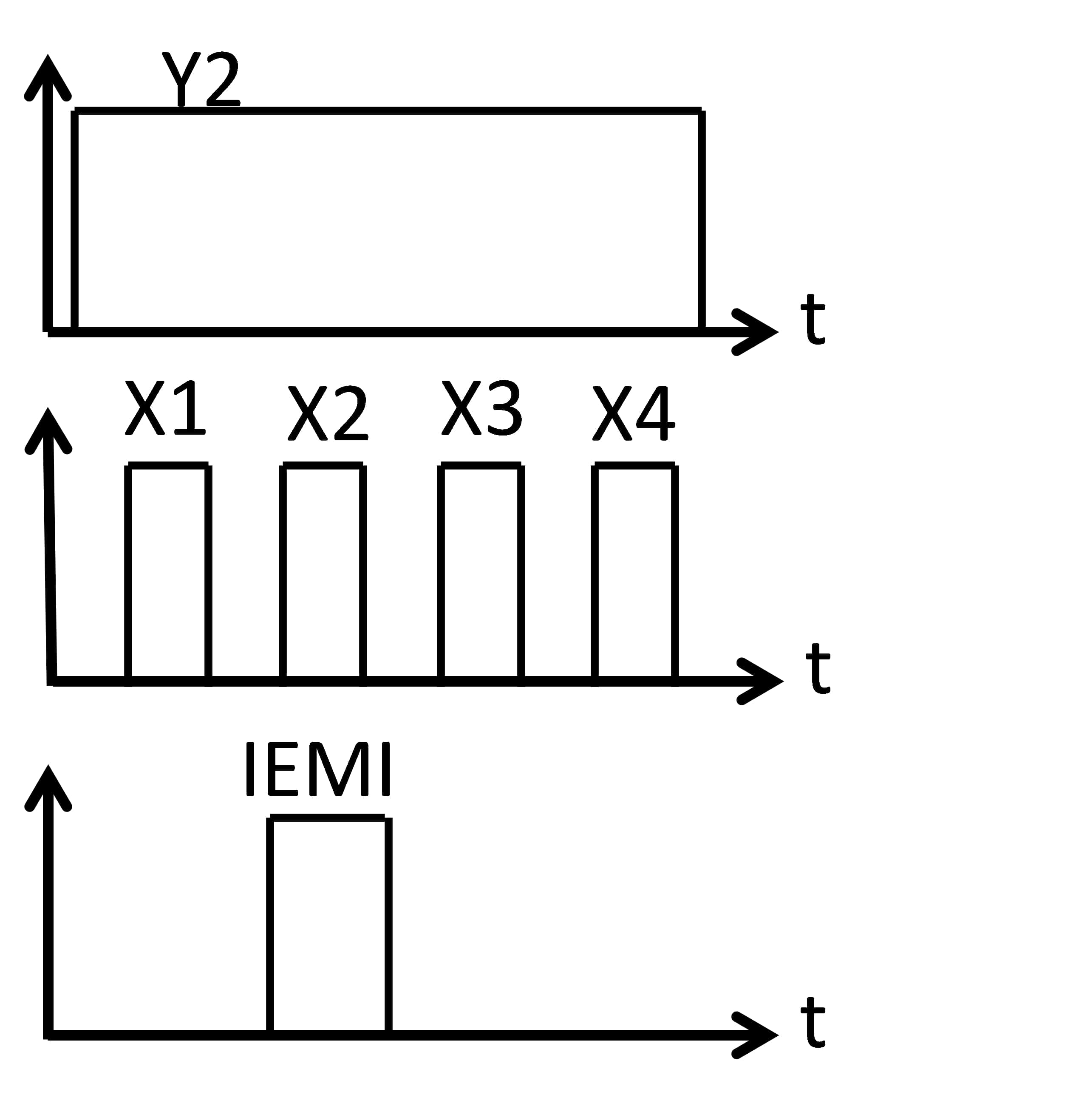}
        \caption{}
        \label{figs:fig19a}
    \end{subfigure}%
    \begin{subfigure}[t]{.65\linewidth}
        \centering
        \includegraphics[width=\linewidth]{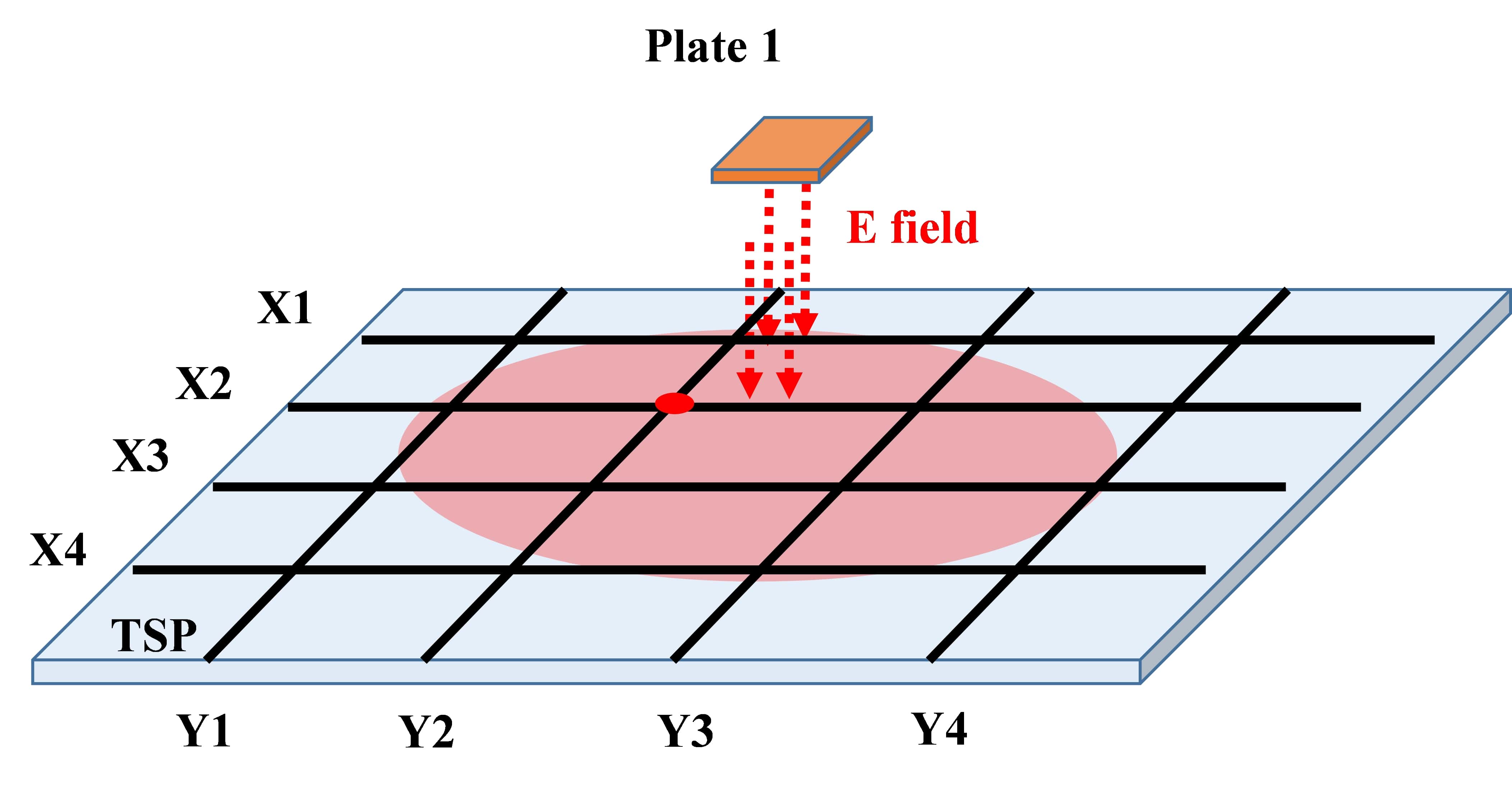}
        \caption{}
        \label{figs:fig19b}
    \end{subfigure}
\caption{Illustration of a precise IEMI attack (a) controller and IEMI signals and (b) ghost touch on a precise location.}
\label{fig:fig19}
\vspace{-.5em}
\end{figure}

In modern touchscreen systems, the electrodes at the touch sensor grid are
scanned by the controller~\cite{barret2010projected}. The controller drives a
single column (TX electrode) and scans every row (RX electrode) as shown in Fig.
\ref{figs:fig19a}. The process is repeated for every column so that the capacitance of all the electrodes can be measured. For example, in Fig.
\ref{figs:fig19a}, column \textit{Y2} is being driven and rows \textit{X1} to
\textit{X4} are being sensed in sequence. When the IEMI attack on the screen
occurs at the moment when a single pair of electrodes is being scanned (see Fig.
\ref{figs:fig19b}), it is possible to generate a ghost touch at that specific
location. A ghost touch will be recognized at \textit{(X2, Y2)} when IEMI occurs while those electrodes are being sensed.

Generating an E field with a small focusing area is challenging. However,
it is possible to generate a ghost touch at a specific location on the
screen without synchronizing with the sense lines if the IEMI signal is
generated with an appropriate antenna using a short pulse. This essentially
mimics a finger touch event. In Section~\ref{sec:poc}, we use two copper
plates which are attached to the front and back of the victim device to
generate a focused small E field. Although such a setup is impractical in
real attack scenarios, we can use the same methodology to design a new
antenna, e.g., using two copper plates right next to each other. In this
design, one copper plate is connected with an excitation signal and the
other is connected to ground. With this configuration, the generated E
field is drawn into the grounded copper plate rather than distributed on
the surface of victim device. In our later experiment section, we show that
our antenna design can be made as small as 4mm x 4mm which provides both
accuracy and high resolution. In section~\ref{sec:new_attack_setup}, we
show how a copper needle antenna can be used on a large touchscreen
device to generate highly accurate ghost touches without the involvement of
ground due to the internally large metal of the device.

\section{Features Affecting IEMI Attack Performance} 
\label{sec:features}

In this section, we evaluate the accuracy and effectiveness of our touchscreen
attack with different touchscreen devices across different manufacturer, size,
operating system, and model. We explore the features affecting IEMI attack
peformance and practicality. In particular, we highlight the success rate and
accuracy of the IEMI attack using different materials and at different
distances. We also demonstrate how to locate the position of the phone and
manage interference between antennas.   

\subsection{Experimental Setup}\label{sec:new_attack_setup}

\begin{figure}[ht]
    \centering
    \includegraphics[width=0.9\linewidth]{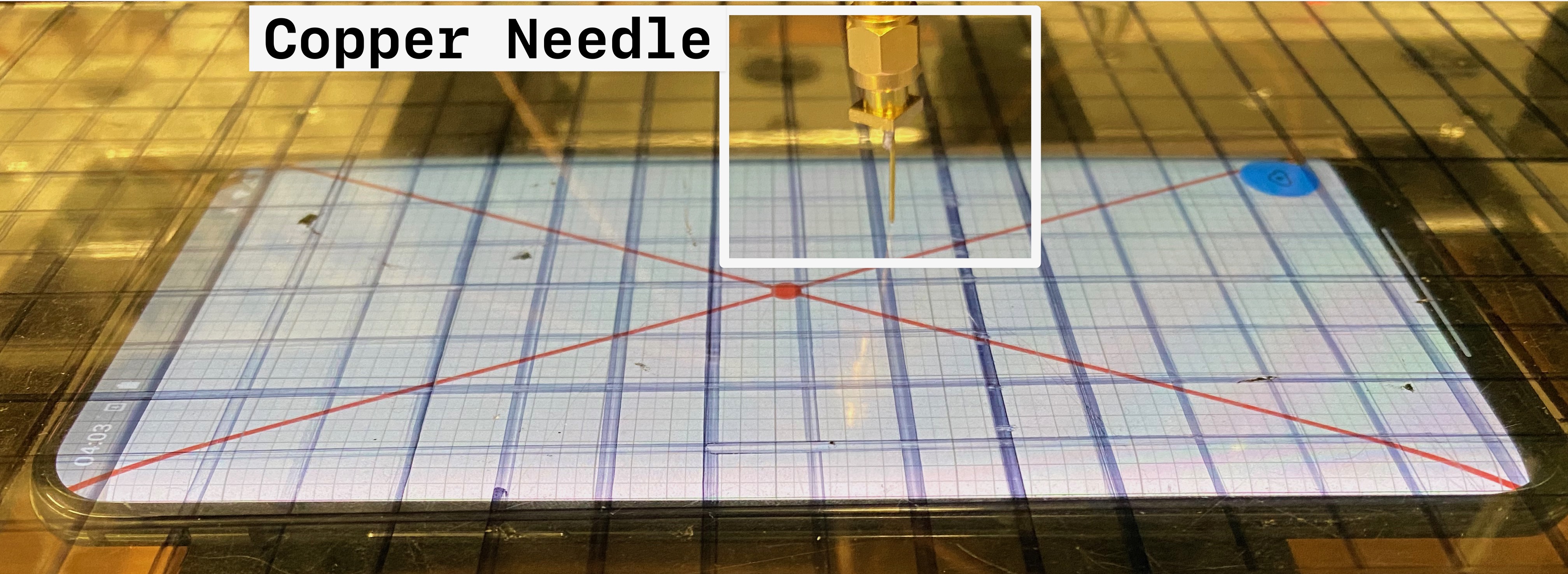}%
    \caption{Copper needle antenna and device under test
    \label{fig:ipadpro_probe}}%
\end{figure}

To evaluate how different factors can influence the generation of a ghost touch,
we conduct experiments using a similar setup as presented in
Section~\ref{sec:poc}, except we add a probe positioning system and single-end
antenna, as shown in Fig.~\ref{fig:ipadpro_probe}. We use standard SMA-to-SMA
coaxial cables which are equipped with a shielding layer to connect the antenna
to the RF amplifier to avoid undesired EM signal emission. It is worth
mentioning that we use copper needles as antennas for our experiments on the
iPad Pro and Surface Pro devices because they provide better resolution due to
the more focused E field at the needle tip. As for the smaller devices tested,
such as iPhone and Android smart phones, we still use the standard copper plates
(4mm x 4mm) antenna setup because it provides a more controllable and small E
field due to the presented ground terminal. We attach the copper plate/copper
needle to standard SMA connectors as the antenna. A separate copper plate is
also used to measure the touchscreen sampling signal for the phone detector
which we will elaborate in Section~\ref{sec:phone_locator}.

\subsection{Experiment Design}

To evaluate the precision and success rate of our touchscreen attack across
different victim devices (Android, iOS, Windows), we designed our own
cross-platform touchscreen gesture collection application with flutter. The
application collects tap, double tap, long press, and swiping gestures on the
touchscreen. It then reports all detected gestures and their associated time and
location to a remote server for subsequent analysis. The application draws a red
dot at the center of the test device for target visualization purposes. The
application also visualizes the detected gestures on the screen along with
coordinates information.

\subsection{Success Rate and Accuracy}

With the reported touch event location and timing, we can perform evaluation
against the collected data to show both the success rate and accuracy of our
attack. During the experiment, we notice that our attack occasionally creates
rare random touch events at distant positions due to the non-ideal E field
spread and interference from nearby equipment. This is shown in
Table~\ref{table:table_material} under the \texttt\textbf{QD (X)} and
\texttt\textbf{QD (Y)} columns, where we choose Quartile Deviation (QD) to
better evaluate how the generated touch events are focused in a small region.
The \texttt\textbf{QD (X)} and \texttt\textbf{QD (Y)} columns represent how
large the generated touch events are distributed along the X axis and Y axis of
a test device with respect to pixels. Another benefit of using Quartile
Deviation instead of Standard Deviation is that we find if the generated touch
event is far away from its intended target, then it will not interfere with the
attack chain by, for example, pressing an incorrect button that is adjacent to
the correct button. As the result, we believe QD is an appropriate metric to
quantify the ``actual attack'' accuracy. From Table~\ref{table:table_material}, we
can tell that our attack performs accurately on the iOS device, especially on
large touchscreen devices. However, we also noticed that our attack often
creates scattered touch events vertically or horizontally. After further
investigation, we believe that although our antenna and signal cable is
specifically chosen to generate a small, focused interference signal, there are
still undesired IEMI signals leaked and the Android test devices are sensitive
enough to recognize them as touch events. Note that the ghost touch occurs every
time we apply IEMI signal on these Android devices so the ghost touch success
rate is 100\% but the accuracy is lower than iOS devices.

\def\halfcheckmark{\tikz\draw[scale=0.22](0,.35) -- (.25,0) -- (0.8,.8) -- (.25,.07) -- cycle (0.8,.02) -- (0.8,0) -- (0.1,.8) -- cycle;}

\begin{table*}[ht]
    \caption{Success Rate and Accuracy of Touchscreen Attack}
    \centering
    \label{table:victim_devices}
    \begin{tabular}{c c c c c c c c}
    \toprule
    \tabhead{Device} & 
    \tabhead{Operating System} & 
    \tabhead{Success} & 
    \tabhead{Freqeuncy} (kHz) &
    \tabhead{Electric Field Strength} (V/m) &
    \tabhead{Success Rate} (s) &
    \tabhead{QD (X)} (s) &
    \tabhead{QD (Y)} (s) \\
    \midrule
    Nexus 5X & Android 8.1.0  & \halfcheckmark & 270  & 1000 & 100$\%$ & 3.5 & 182.5 \\
    Google Pixel 2 & Android 10  & \halfcheckmark & 230  & 1000  & 100$\%$ & 10.0 & 149.5 \\
    OnePlus 7 Pro & Android 11 & \halfcheckmark  & 295  & 800  & 100$\%$ & 196.5 & 3.0 \\
    iPhone SE & iOS 12.0 & \checkmark  & 95  & 1500  & 57$\%$ & 10.5 & 6.0 \\
    iPhone 6 & iOS 12.2 & \checkmark  & 98  & 1500  & 86$\%$ & 14.0 & 10.0 \\
    iPhone 11 Pro & iOS 14.7.1 & \checkmark  & 120  & 1500  & 77$\%$ & 4.5 & 8.5 \\
    Surface Pro 7 & Windows 10 Pro 2004 & \checkmark  & 220  & 1200  & 88.3$\%$ & 12.5 & 7.5 \\
    iPad Pro & iPadOS 14.7.1 & \checkmark  & 270  & 1500  & 100$\%$ & 1.0 & 0.5 \\
    \bottomrule
    \end{tabular}
\end{table*}

\subsection{Table Material}

As we aforementioned in Section~\ref{sec:poc}, the dielectric constant of the table
material impacts our attack. To evaluate the performance of our attack using different common table materials, we choose five typical table top samples (solid wood, acrylic, marble, medium density fiberboard/MDF, copper) as the
insulation material between antenna and victim device and repeat our experiment.
We conduct the experiment with acrylic sheet and our probe positioning system
first and then swap the table top sample so that we can still calculate the
statistical dispersion for non-transparent table material. The thickness of
these table material samples are all 10mm. Table~\ref{table:table_material}
shows that when non-metal table materials are used, our attack can achieve similar performance with respect to success rate and dispersion. However, the metal table material does not allow us to perform a valid attack due to its high conductivity. 

\begin{table}[ht!]
    \caption{Touchscreen Attack with Different Table Materials}
    \label{table:table_material}
    \centering
    \begin{tabular*}{\columnwidth}{ccccc}
        \toprule
        \tabhead{Material} &
        \tabhead{Dielectric Constant} & 
        \tabhead{Success Rate} &
        \tabhead{QD (X)} &
        \tabhead{QD (Y)} \\
        \midrule
        acrylic & 2.7 - 4.0  & 100\% & 1.0  & 0.5 \\
        marble & 3.5 - 5.6  & 76\% & 2.6  & 1.0  \\
        solidwood & 1.2 - 5  & 90\% & 1.6  & 1.4  \\
        MDF & 3.5 - 4 & 100\% & 1.0  & 1.0  \\
        copper & \ding{55}  & \ding{55} & \ding{55}  & \ding{55}  \\
        \bottomrule
    \end{tabular*}
\end{table}

\subsection{Table Thickness}
\label{sec:table_thickness}

\begin{figure}[ht]
    \centering
    \includegraphics[width=0.98\linewidth]{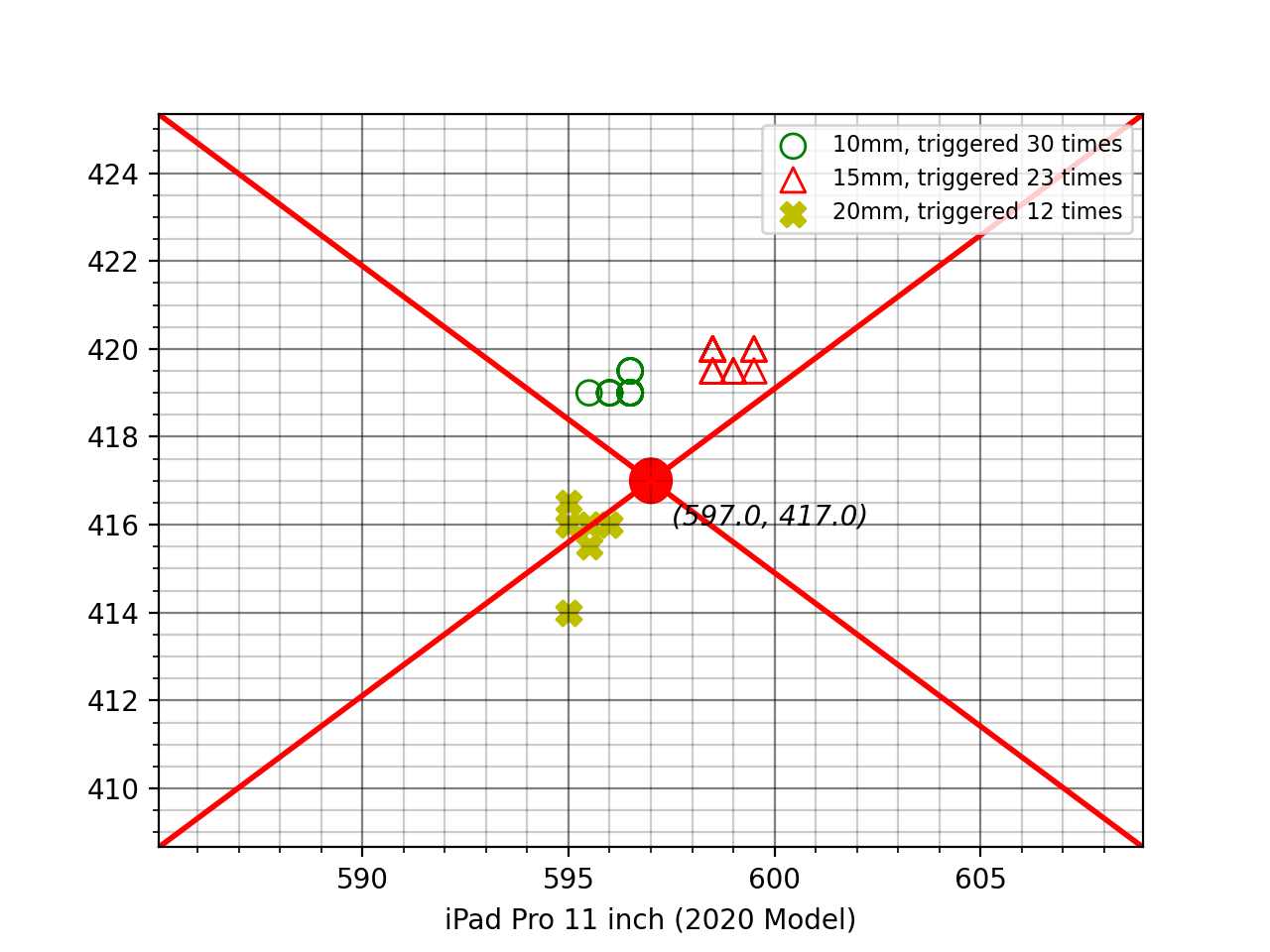}%
    \caption{Generated touch event on iPad Pro with different table thickness.
    \label{fig:ipad_pro_attack}}%
\end{figure}

To understand the practicality of our attack, we also evaluate it with
respect to success rate and accuracy using different thicknesses of table
material. We set the signal generator to sweep mode and each sweep period is set
to 1 second, such that the correct interference frequency will be generated every second. The total time of signal generator output lasts 30 seconds. We use
our own application to record how many touch events are generated during the
test period and where/when they are generated. Using an iPad Pro and acrylic
sheets, we conduct the experiments when the thickness of the acrylic sheets is 10mm,
15mm, 20mm. As we can see in Fig.~\ref{fig:ipad_pro_attack}, the success rate
of our attack is up to 100\% when the table thickness is 10mm. The success rate
decreases to 76\% when the table thickness is 15mm. The success rate eventually
drops to 40\% when the table thickness is 20mm. In real life, the common table thickness is only 1/2 inch or 5/8 inch based on IKEA~\cite{ikea}, Office Depot~\cite{office-depot} and Wayfair~\cite{wayfair}. Our effective attack distance, 20mm, is larger than the common tabletop thickness.

\subsection{Interference Between Antennas}
In our experiments, we design and use an antenna array to generate multiple touch events at different
locations. However, if we need sequential touch events, only one antenna will be
applied with an excitation signal at a certain time and other antennas should be kept
as either grounded or floated. However, two antennas that are physically close
with each other can easily couple with each other and create undesired touch
events at random locations and times. To overcome this issue, we employed
isolated and shielded signal cables and antennas. All the signal cables that are used to drive the antenna array are standard SMA-to-SMA shielded cables in order to avoid coupling between each other. Furthermore, copper tape is used to cover the antennas to insulate the generated EM field into a  small region as shown in Figure~\ref{fig:ipadpro_probe}.

\section{Practicalities of Touchscreen attack }
\label{sec:practicalities}
In this section, we discuss how to utilize the proposed IEMI attack in real
attack scenarios. To perform a practical attack, the attacker has three major
obstacles to overcome, the design of an IEMI antenna, knowledge of the victim
device's location, and knowledge of a successfully injected touch event. We
address all three obstacles by building an antenna array, phone locator, and
touch event detector respectively. 

\subsection{Design of an IEMI Antenna}

In previous sections, we show how to inject simple tap, long hold, and any
direction sweep gestures on touchscreens with a single needle IEMI antenna. The
injected touch gestures are located directly in the path of the IEMI antenna.
Under a practical scenario, however, the touchscreen device can be randomly
placed on the tabletop. A single needle IEMI antenna is therefore insufficient
to inject a touch event if not placed directly in its path. We consider two
solutions to address this issue. First, the attacker can implement a mechanical
system to maneuver the single needle IEMI antenna into the desired location of
the victim touchscreen device, then perform an IEMI attack. The attacker can then
operate the IEMI antenna to perform complicated drawing gestures by continuously
generating the interference signal to meet the attack requirement. While
possible, we consider this a less-than-ideal solution due to both the size and
noise of the mechanical infrastructure required to freely move a single needle
IEMI antenna under a tabletop without being detected. This option would therefore require significant effort and cost to ensure a stealthy design. We therefore opt for implementing a static antenna array to
reduce the associated engineering and practical issues mentioned above. A modular antenna array allows us to configure the way it is
attached, so that we can increase the density of IEMI antennas for a smaller
target device without changing the hardware design. In addition to the antenna
array, we implement an IEMI channel controller that can independently control up
to 64 IEMI antennas using programmable reed relays. The size of the designed
IEMI channel controller and antenna array are smaller enough to squeeze into a
shoe box. The needles of the antenna array are inserted into foam to support and
protect the fragile hardware. The size of the array is 24cm x 17cm,
and the distances between the antennas vary between 2cm and 7mm to meet the
density requirements for different sizes of target touchscreen devices. 

\subsection{The Screen Locators}
\label{sec:phone_locator}
As we have mentioned in Section~\ref{subsec:capactive_touchscreen}, a
touchscreen sensing system consists of a grid of TX and RX electrodes. The TX
electrodes generate varied excitation signals on different lines while the
intersecting RX electrodes sense the physical variations to determine the touch
points. Our experiments found that antennas placed near the screen can easily
pick up these TX signals. Such signals contain patterns that can tell us at
which TX lines the antennas are pointing. Besides, when an antenna is placed
perpendicular to the screen, only the pointed TX electrode produces the
strongest signals, while nearby electrodes have little impact on the received
signals. Hence, the signal received by an antenna can be used to identify the
pointed-at location with high spatial resolution. For example, a significant
signal strength degradation can be observed when two antennas are placed on both
sides of a screen boundary. This feature allows us to accurately detect the
screen boundary location with an error of less than 1 cm.

\begin{figure}[htbp!]
	\centering
	\subfloat[Sequential driven TX signals\label{fig:drivingMethod_a}]{\includegraphics[width=0.5\columnwidth]{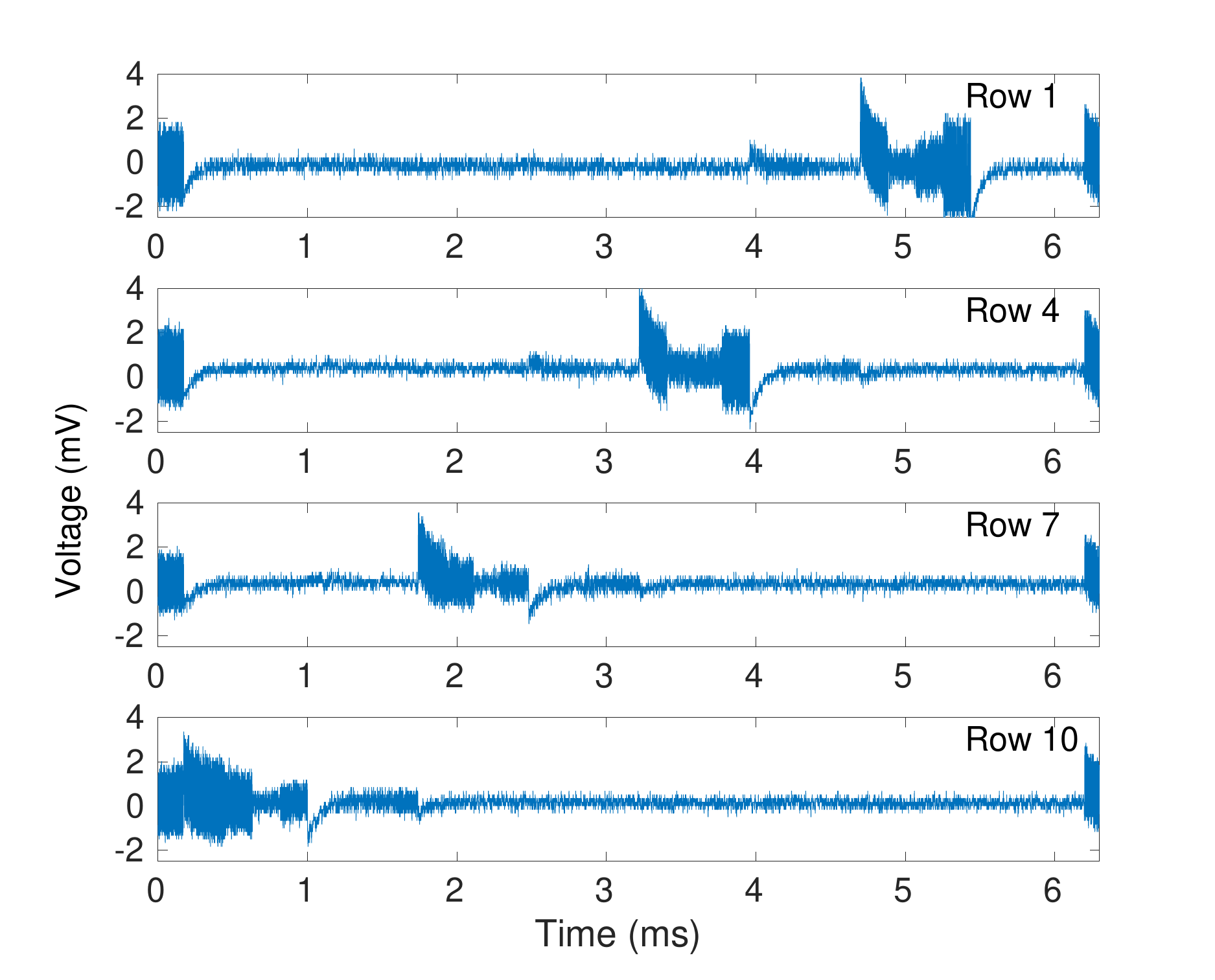}}	
	\subfloat[Parallel driven TX signals \label{fig:drivingMethod_b}]{\includegraphics[width=0.5\columnwidth]{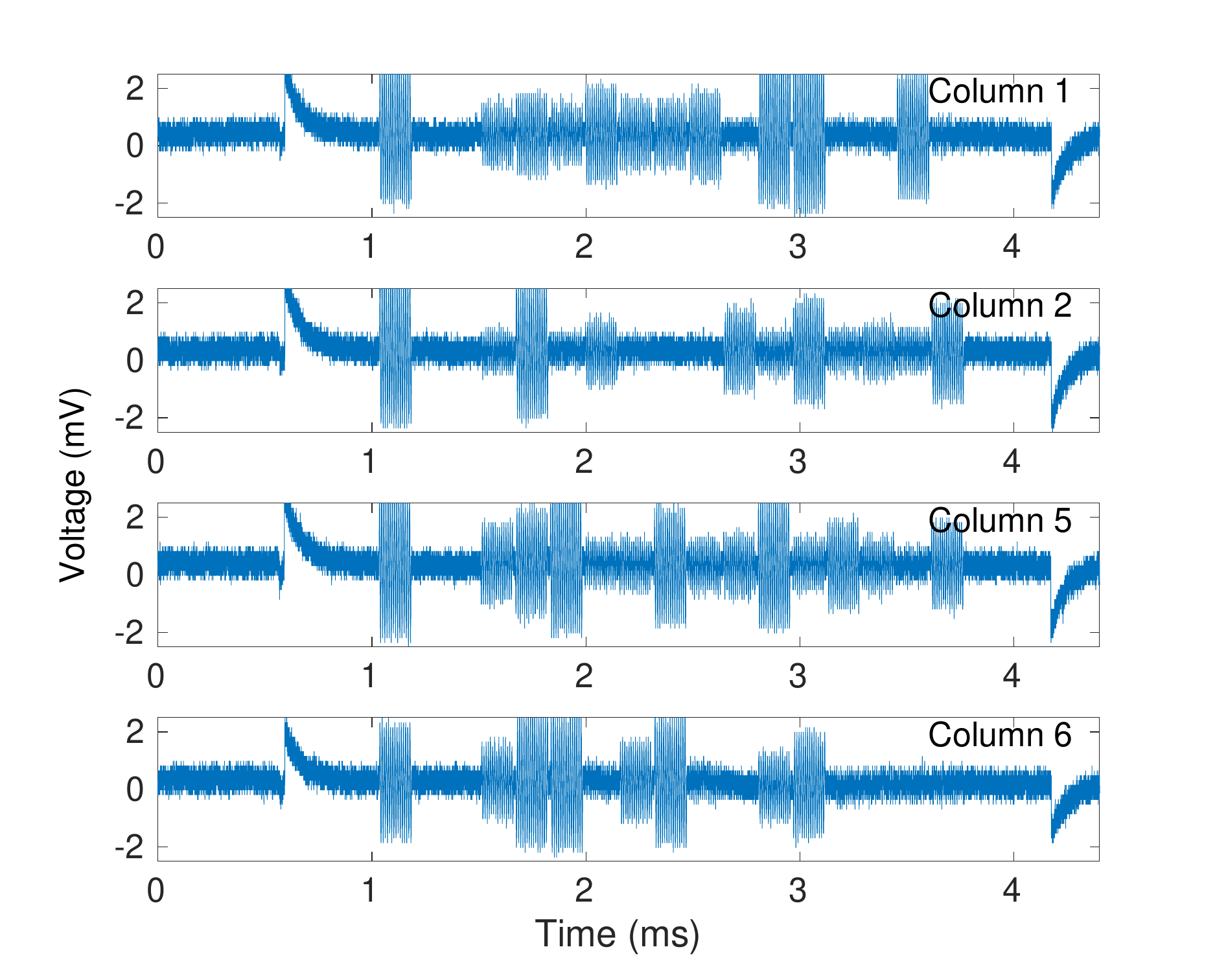}}
	\caption{TX signals on screens with different driving methods}
	\label{fig:drivingMethods}
\end{figure}

Various driving methods can be used to generate the TX signals. Among all
examined devices, we observed two methods being used. The sequential driving
method (SDM) is usually implemented to excite the electrodes in turn. As a
result, the electrode location can be identified by checking when a TX signal
appears. Fig.~\ref{fig:drivingMethod_a} shows EM traces collected on four
different rows of a Google Pixel 2. We can observe the linear relationship between
the rows and the appearing time of TX signals. The orientation and location for
this kind of screen can be quickly recovered using a simple linear function.
Besides the sequential driving method, we found the parallel driving method
(PDM) to be a more frequently implemented technique on most of the latest
devices, which uses orthogonal codes to drive all TX signals concurrently.
Fig.~\ref{fig:drivingMethod_b} shows EM traces collected on four different
columns of an iPhone 11 Pro. As we can see, instead of generating signals with the
same patterns sequentially, different electrodes produce signals with varied
patterns simultaneously. In this case, recovering the location information is
more challenging because of the less straightforward correlations between
signals and screen locations. However, we can still successfully recover the
screen location information using these TX signals with the technique described
below.

Our technique consists of three steps: feature extraction, classifier training,
and location prediction. As shown in Fig.~\ref{fig:drivingMethod_b}, the
boundaries between two code bits can be identified, which allows us to segment
the signals corresponding to each code bit. For each segment, we can compute
descriptive features for a code bit, which can be the phase, the magnitude, or
the frequency, depending on the specific encoding schemes used by the screen.
Then, we can derive a feature vector for each TX signal by concatenating these
features. Afterward, we can train a classifier with enough feature vector and
location pairs. This classifier can identify the screen location using the
signal collected at an unknown location.

We can identify different TX electrodes in different lines using this technique,
but we can not distinguish different locations on the same TX electrode. Expressed differently, for any antenna with a known antenna coordinate
$(x_\texttt{antenna}, y_\texttt{antenna})$, we can obtain a single dimension
screen coordinate, which may be $x_\texttt{screen}$ or
$y_\texttt{screen}$. To determine the other dimension, we also need to know at
least one antenna coordinate mapped to the screen boundary to tell us the
unknown dimension. As mentioned above, the screen boundary can be accurately
located by looking for significant signal strength degradation between two
adjacent antennas. With enough antenna coordinate and screen coordinate pairs,
we can derive the mapping between them. The mapping between $(x_\texttt{screen},
y_\texttt{screen})$ and $(x_\texttt{antenna}, y_\texttt{antenna})$ can be seen
as a rotation followed by a translation as described in
Equation~\ref{eq:transformation}, where $\theta$ represents the rotation while
$x_t$ and $y_t$ represent the translation. After solving this equation, we can
use this transformation matrix to select the closest antenna to inject the error
for any target screen location.
\begin{equation}\label{eq:transformation}
	\begin{bmatrix}
	x_\texttt{screen} \\
	y_\texttt{screen} \\
	1
	\end{bmatrix}
	=
	\begin{bmatrix}
	cos(\theta) & -sin(\theta) & x_t\\
	sin(\theta) & cos(\theta)  & y_t \\
	0           & 0            & 1 \\
	\end{bmatrix} 
	\begin{bmatrix}
	x_\texttt{antenna} \\
	y_\texttt{antenna} \\ 
	1
	\end{bmatrix}
\end{equation}

\begin{figure}[htpb!]
	\centering
	\subfloat[\centering Screen location detected using 7 antennas \label{fig:iPadLocDetect_a}]{
	\includegraphics[width=0.45\columnwidth]{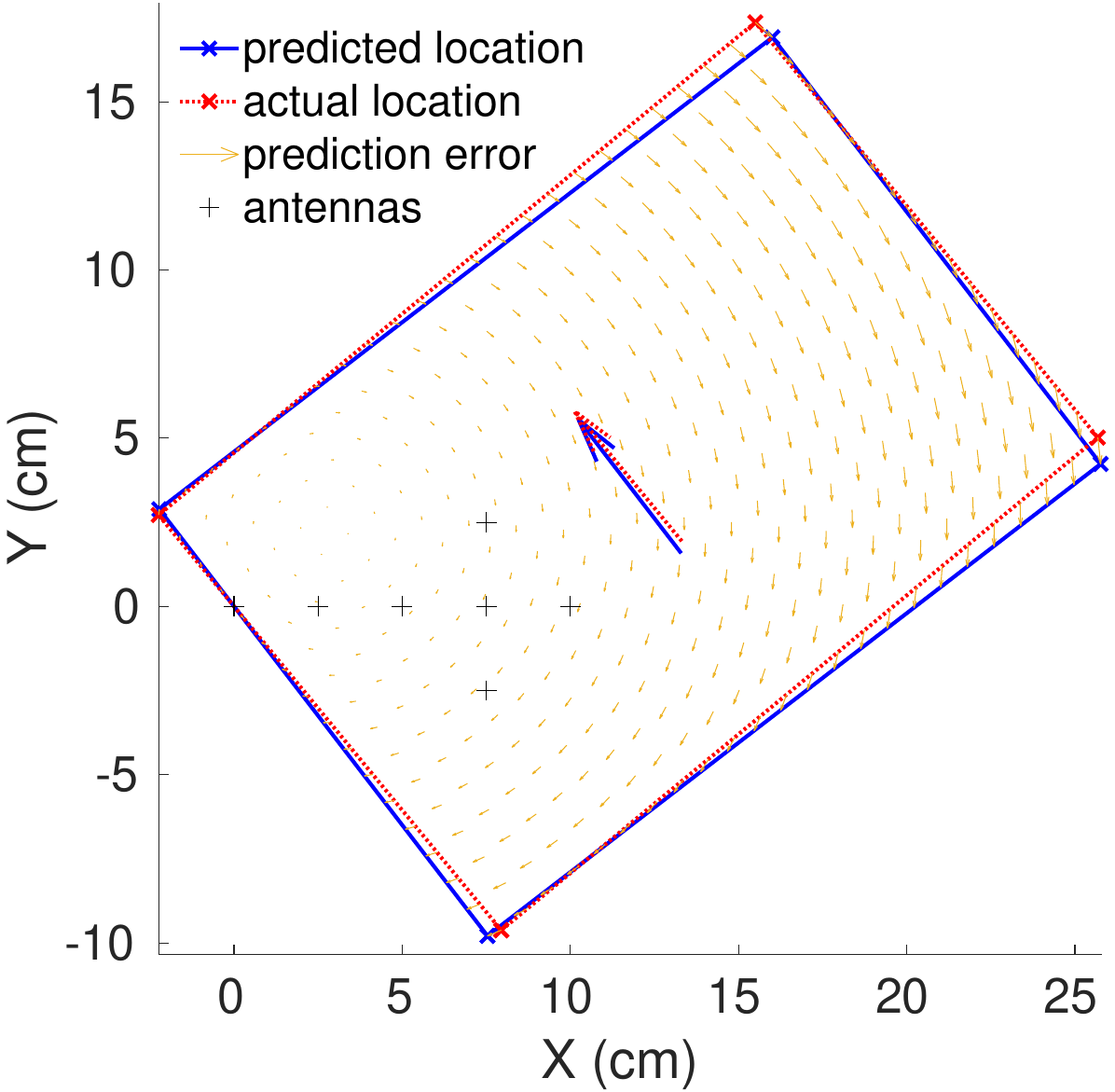}}
	\subfloat[\centering Screen location detected using 12 antennas \label{fig:iPadLocDetect_b}]{
	\includegraphics[width=0.45\columnwidth]{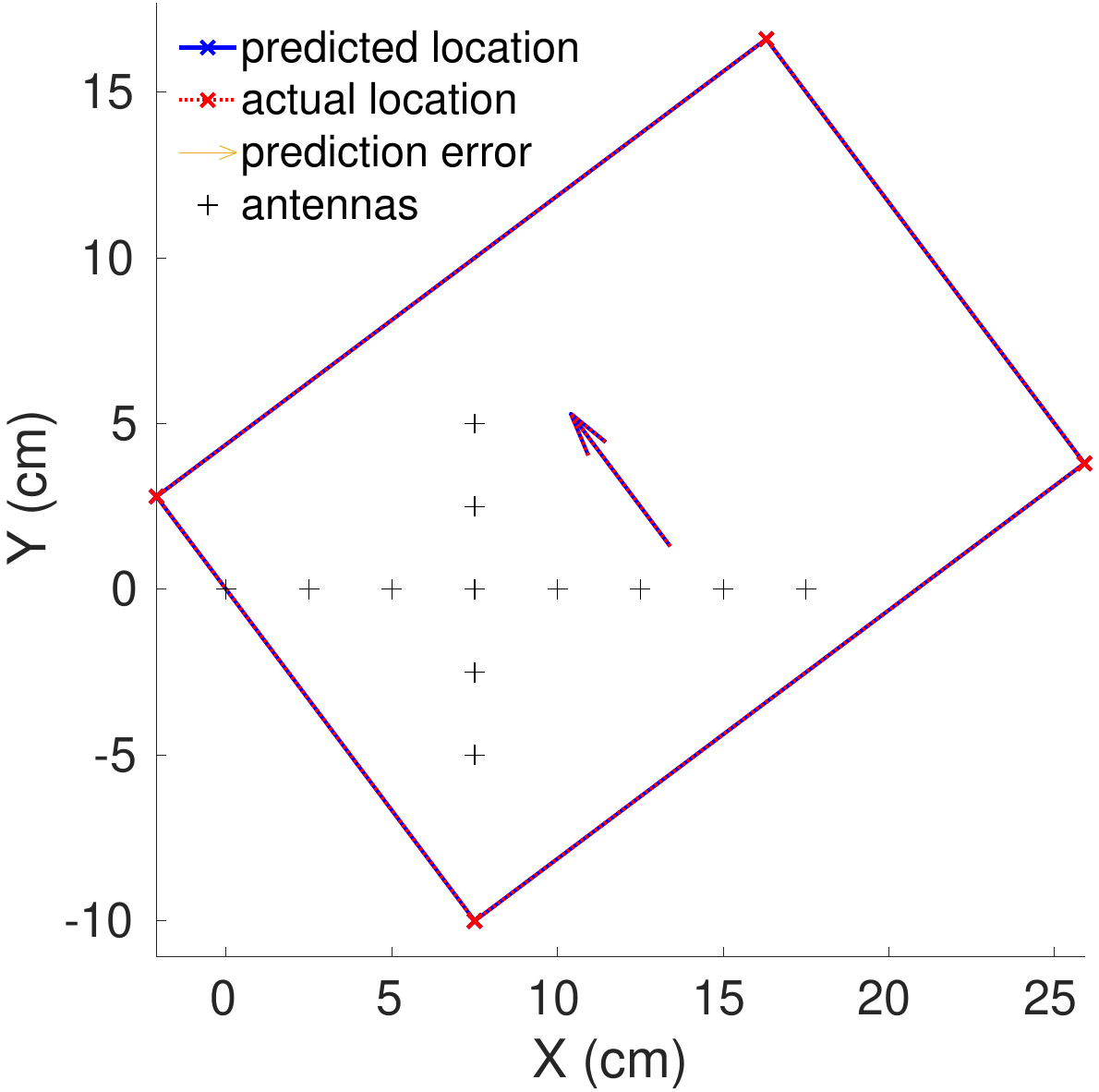}}
	\caption{Screen location detection results of iPad Pro}
	\label{fig:iPadLocDetect}
\end{figure}

To better demonstrate how the screen locator works, we use an iPad Pro as an
example. From a TX signal on the iPad Pro, we can obtain a feature vector with 48
feature values using the magnitude of sinusoidal signals in each segment, which
is correlated to the row number on screen. Signals are collected from the bottom
row to the top row with a step of 1cm. On each row, signals are collected at
12 different columns. These signals are used to train a k-nearest neighbors (KNN)
classifier. In the evaluations, we first use signals collected from 7 antennas
in a small area to detect the location and orientation of the tested iPad Pro.
Fig.~\ref{fig:iPadLocDetect_a} shows the detection results. The predicted
location is pretty close to the actual location, with maximum prediction error
being 0.8cm. Furthermore, if we use 5 more antennas to collect signals in a larger area, the prediction result matches perfectly with the actual
location.

We tested our screen locator on 5 devices listed in Table~\ref{table:locator_results}.
We list the driving methods used by these devices, the sample rate we use to collect the data, the average prediction error, and the average computation time.
Note that for screens using SDM, the location is computed using the time stamp read from an oscilloscope.

\begin{table}[ht!]
    \caption{Screen Location Detection Results}
    \label{table:locator_results}
    \centering
    \begin{tabular*}{\columnwidth}{ccccc}
        \toprule
        \tabhead{Device} &
        \hspace{-1em} \tabhead{Driving Method} &
        \tabhead{Sample Rate} &
        \tabhead{Error} & 
        \tabhead{Time} \\
        \midrule
        Nexus 5X & SDM & 50MSa/s & 0.42 cm & N/A \\
        Google Pixel 2 &  SDM & 50MSa/s & 0.51 cm&  N/A \\
        iPhone 11 Pro & PDM & 1MSa/s & 0.3 cm &  0.08s \\
        OnePlus 7 Pro & PDM & 2MSa/s & 0.06 cm & 0.14s \\
        iPad Pro & PDM & 1MSa/s & 0.18 cm & 0.17s \\ 
        \bottomrule
    \end{tabular*}
\end{table}

\subsection{The Touch Event Detectors}
\begin{figure}[ht]
    \centering
    \includegraphics[width=0.9\linewidth]{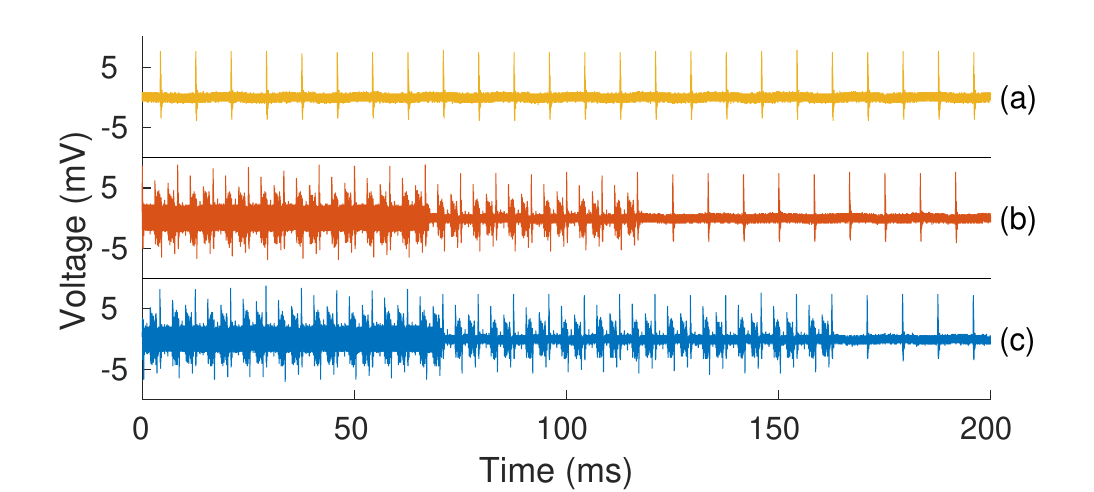}%
    \caption{Emission signal from iPad Pro (a) reduced scan. (b) failed IEMI attack. (c) successful IEMI attack.
    \label{fig:ipad_scan}}%
\end{figure}

To perform an attack which requires several touch events to complete, it is
important to know whether the current touch event injection is successful before
proceeding to inject the next touch event at a different location. In certain
cases injection of a successful touch event may take more time than expected.
As introduced in Section~\ref{sec:related}, there are multiple techniques to
detect the current screen content out of sight. However, these techniques can be
difficult to use without significant effort. In our work, instead of detecting
if we have altered the screen content as desired, we detect if our last touch
event injection was successfully applied on the screen. The key behind such
detection is the active scanning mechanism used by modern touchscreen
controllers~\cite{microchip_2005}. To achieve balance between the
power efficiency and scanning accuracy, touchscreen controllers perform reduced scanning to preserve the power. Once a touch event is detected on the
touchscreen, the controller changes the scanning mode from reduced scan to full
scan to measure the touched location more accurately. If there are no more
touch events detected, the controller switches back to reduced scan mode
automatically. Although we do not have a datasheet for a commercial touchscreen
controller, using our IEMI antenna we observed similar behavior on all tested touchscreen devices. More importantly, if the touch event is successfully
injected on a target device and recognized by the operating system, the
touchscreen controller takes a longer time to switch back to reduced scan mode. As
shown in Figure~\ref{fig:ipad_scan}a, the iPad Pro emits a sparse scanning signal with 120Hz frequency when no finger or IEMI signal is present. Figure~\ref{fig:ipad_scan}b shows how the touchscreen switches from full scan mode back to reduced scan mode after we turn
off our IEMI signal. We can also see the touchscreen recognizes our IEMI
signal as a touch event but eliminates it due to the wrong interference frequency.
In Figure~\ref{fig:ipad_scan}c, we apply a correct IEMI signal and successfully trigger a
touch event on screen. The time that the controller takes to switch back to
reduced scan mode is discernibly longer compared to the previous
experiment. Such phenomena is stable and is exhibited on all our tested devices. Using this technique, we examine the collected touchscreen emission signal
right before we turn off the IEMI attack and detect if any touch event was injected in the previous attempt. Our experimental results show that this approach
works every time on our three main test devices (iPad Pro, iPhone 11 Pro and
Oneplus 7 Pro). The touch event detector is implemented as a dedicated IEMI antenna which connects to an oscilloscope. 

\section{Evaluation of Practical Attacks}
\label{sec:evaluation}
\subsection{The Attack Setup}
\begin{figure}[ht]
    \centering
    \includegraphics[width=0.9\linewidth]{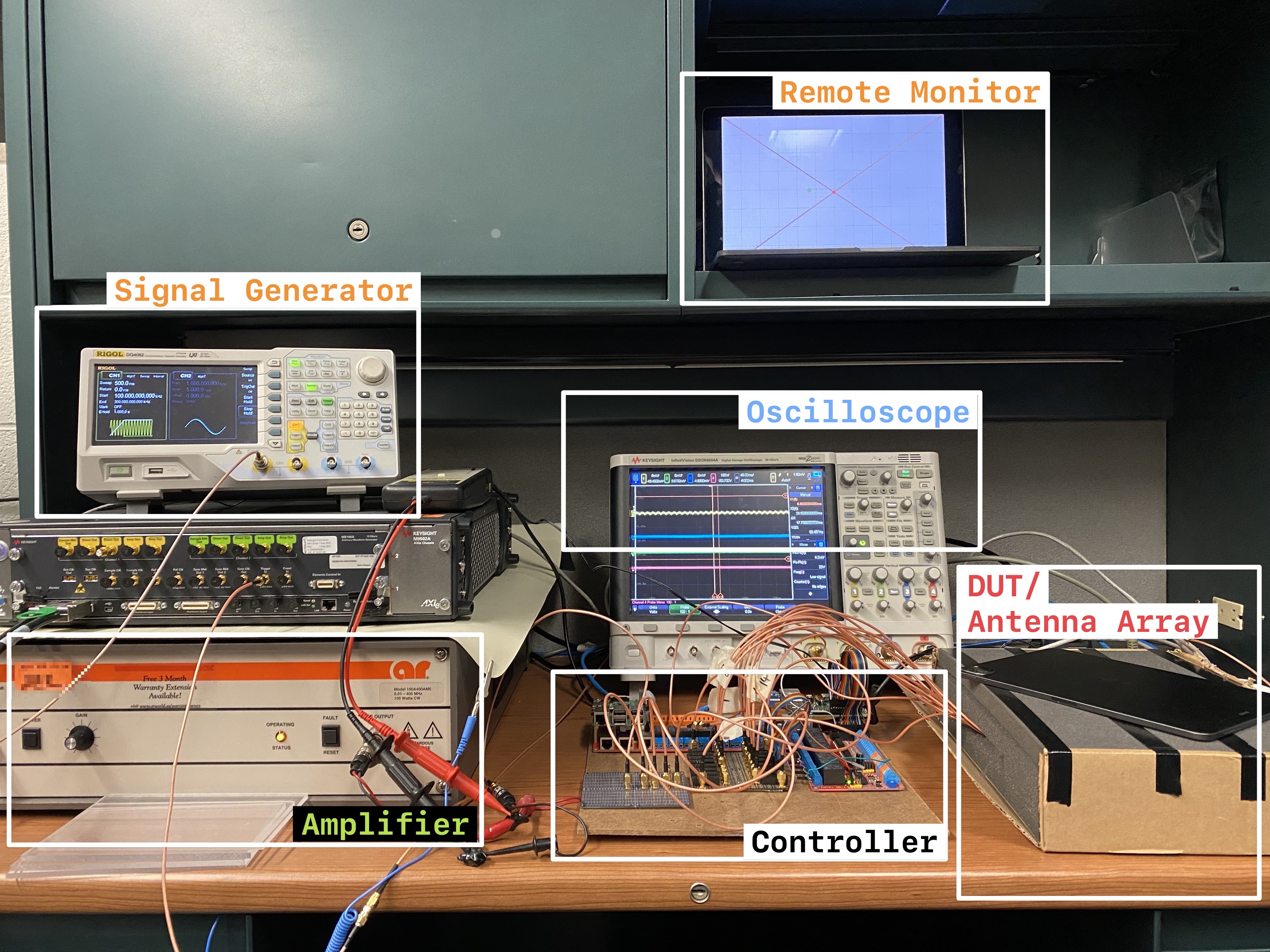}%
    \caption{Attack setup for precision evaluation
    \label{fig:attack_setup_percision}}%
\end{figure}

With our antenna array, phone locator and touch event detector in place as shown
in Figure~\ref{fig:attack_setup_percision}, we are ready to conduct an actual
attack that mimics practical scenarios. We tape our antenna array under the
left-bottom corner of an experimental bench made of MDF with a table thickness
of 15mm. A laptop is placed at the left side of the table outside of the
detect/attack range of our antenna array. During the experiment, we ask ``the
victim'', who has no prior knowledge of the exact location of our antenna array,
to sit in front of our experimental bench and put our unlocked test target
device facing down. We then use our phone locator to infer the current position
and orientation of our target device, perform the attack vectors and monitor the
injected touch events. Note that we do not ask ``the victim'' to use their own
devices as we may alter or leak private content of the target device during the
experiments. 

\begin{figure}[ht]
    \centering
    \begin{subfigure}[t]{.47\linewidth}
        \centering
        \includegraphics[width=\linewidth]{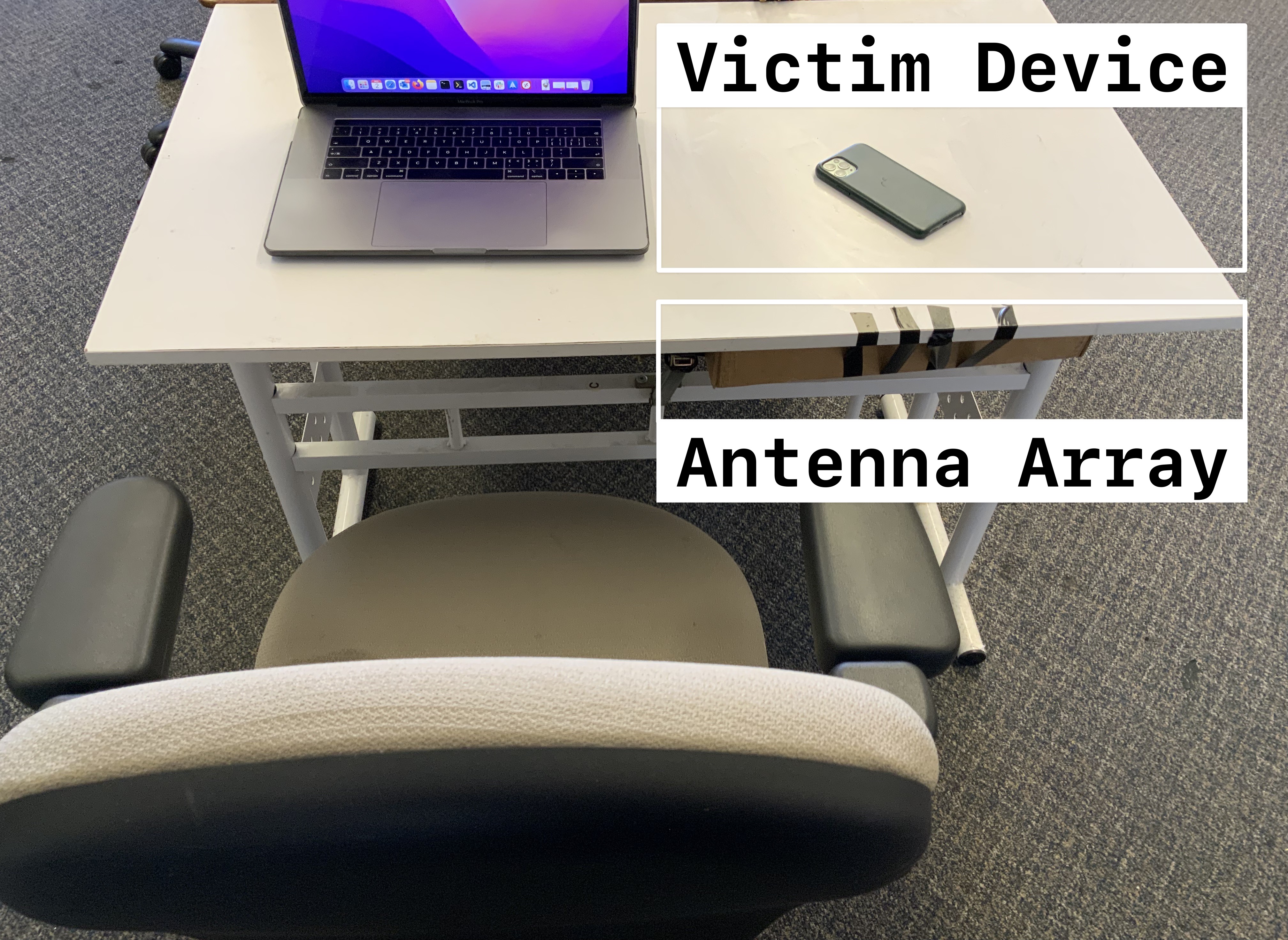}
        \caption{}
        \label{fig:attack_setup_realbench_a}
    \end{subfigure}%
    \hspace{1em}%
    \begin{subfigure}[t]{.47\linewidth}
        \centering
        \includegraphics[width=\linewidth]{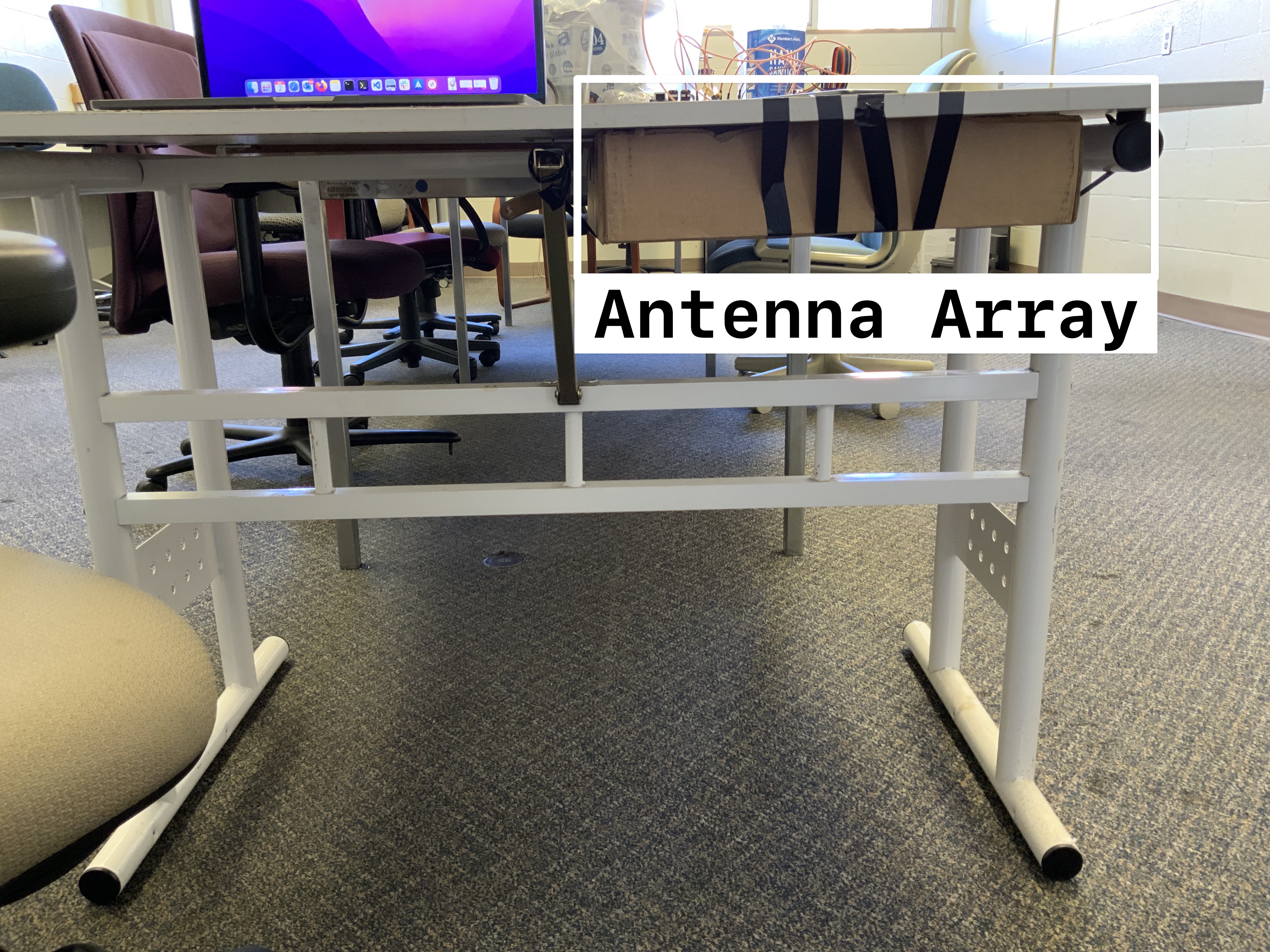}
        \caption{}
        \label{fig:attack_setup_realbench_b}
    \end{subfigure}
\caption{Attack setup with actual table (a) attack setup on the table (b) antenna array attached to the table. }
\label{fig:attack_setup_realbench}
\vspace{-.5em}
\end{figure}

\subsection{Attack Evaluation}

To evaluate the setup in a practical scenario, we choose three different
touchscreen devices as our target devices: 1) an iPad Pro 2020; 2) an iPhone 11
Pro; and 3) a Oneplus 7 Pro. These three devices are pre-installed with our
touch event detection application and remotely mirror their current display onto
another monitor. Note that this application is only installed to better
illustrate the injected touch events during the experiment. Attackers can perform a
similar attack without installing the application ahead-of-time. The test device is
unlocked and randomly placed on our antenna array with different angles and
orientations as described above. We first use the antenna array to capture and
analyze the emitted signal from the target device to predict its current
position and orientation. We have found in our experiments that our phone
locator program typically needs 4 antennas at different locations to infer the
phone location within 3 seconds with a sampling rate of 1M/s. Once we have the
precise location of the target device, we switch the antenna array from monitor
mode to attack mode by switching the corresponding relays. We choose the
appropriate interference frequency and amplitudes based on the target phone
model. We then use our attack setup to launch two different type of attacks
against the touchscreen devices under test using either a precise touch event
injection or sequence of touch events at different locations as needed. 

\noindent\textbf{Leveraging Siri on iOS devices} Installing unauthorized
applications on an iOS device can be difficult due to strict iOS application
distribution. Instead, we leverage our touch event injection attack to abuse
Apple's accessory discovery mechanism to perform data exfiltration. An iOS
device automatically finds nearby unpaired Apple accessories, such as Airpods
headphones. Once these devices are found, a notification pops up and asks the
user if the device should pair and connect. The notification issues a
\texttt{Connect} request that prompts the user to grant access. To connect
with the device the user only needs to tap the \texttt{Connect} button without
further action. Once connected, the user can directly uses the Airpods to wake
up and interact with Siri, the voice assistance on Apple devices. The
\texttt{Connect} request notification is always displayed at a fixed location.
In our experiments, we find the size of the \texttt{Connect} button is
approximately 5.5 cm by 1 cm. The confirmation button occupies roughly 2/3 of
the screen width on an iPhone Pro 11 which makes it easier to attack. On the
contrary, the size of this button on an iPad Pro is much smaller compared to the
size of the screen. However, our attack is still feasible on the iPad Pro due to
its accuracy (see Section~\ref{sec:features} and
Table~\ref{table:victim_devices}). We first conduct an experiment to validate
the possibility of such an attack on a randomly placed iPhone 11 Pro and iPad
Pro 2020 using unpaired Airpods. After successfully pairing with the Airpods we
wake up Siri to read out the new messages of the victim devices. To further
evaluate the success rate of our attack on iOS devices, we use our touch event
application to draw a square space of the same size as the confirmation button.
We randomly place the victim device on our antenna array and repeat the process
of sensing/attack/detection and then evaluate if the injected touch events falls
into the intended region. Our attack works 6 out of 10 times on iPad Pro with an
no more than 12 seconds of attack time and works 9 out of 10 times on an iPhone 11
Pro with no more than 9 seconds of attack time. The random placement of test
devices outside the range of our antenna array are not included in the metric
calculation. During the experimentation, we find that the main point of failure
for an attack on an iPad Pro is that the distance between our IEMI antennas is
too large to have at least one IEMI antenna placed on top of the confirmation
button. The current configuration of number of IEMI antennas and the distance
between IEMI antennas is a tradeoff between antenna array coverage and antenna
density that should be selected based on the target device screen size.

\noindent\textbf{Installing malicious applications on Android devices} To attack
Android based touchscreen devices, we use our IEMI to inject multiple touch
events at different screen locations. More specifically, we assume the attacker
knows the phone number of the victim device and sends it a message which
contains the link of a malicious application. To install the malicious
application, we need to generate 5 distinct touch events in sequence at
different locations, including a tap on the notification of new message (1 large
clickable area), choose action for link (2 buttons in a row, open link/copy
text), allow saving the APK file (2 adjacent buttons), install the APK file
after downloading (1 button), and finally open the APK after installation (2
adjacent buttons). We use a Oneplus 7 Pro to evaluate this attack. We first
measure the location and orientation of the victim device. We then initiate the
attack by sending a message containing the download link of designated
application. Once the message is sent, we use one IEMI antenna that points to
the middle of the screen and two IEMI antennas at the bottom part of the screen
to inject the five touch events in sequence. Each individual touch event is
evaluated with our touch event detector before moving on to the next touch
event. We conducted 10 experiments with different cellphone locations. We
achieved three successful attacks with our setup. Using the mirrored display, we
find that most of the failed attempts were due to incorrectly inducing a touch
event on adjacent buttons. For example, the injected touch event incorrectly
presses the \texttt{CANCEL} button and causes the entire attack to immediately
fail. We believe a better designed IEMI antenna would allow us to focus the
generated E field on a smaller attack area, thereby making our attack more
robust. 

\begin{figure}[ht]
    \centering
    \begin{subfigure}[t]{.47\linewidth}
        \centering
        \includegraphics[width=\linewidth]{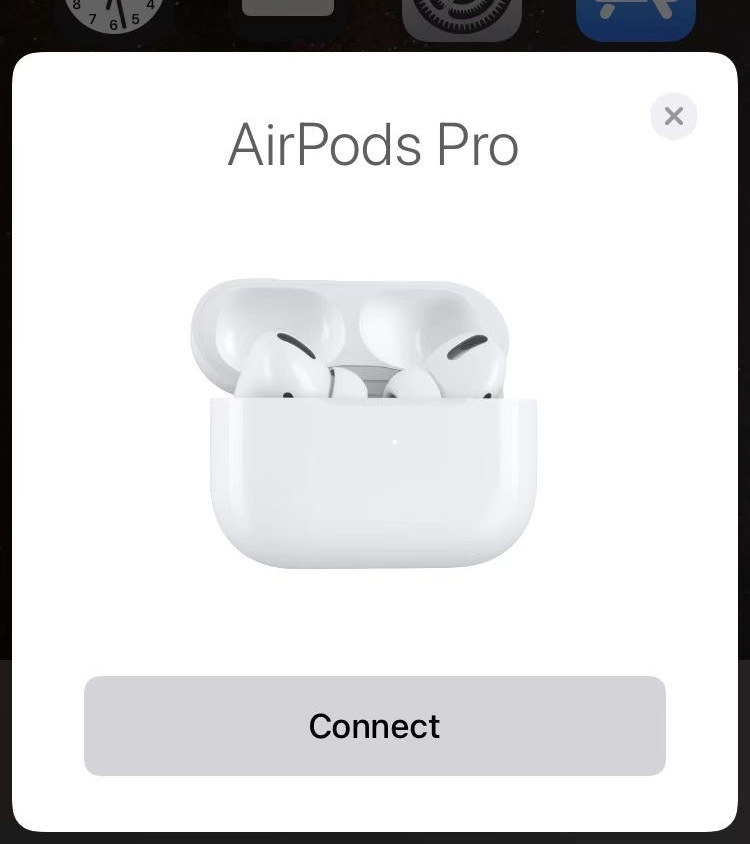}
        \caption{}
        \label{fig:attack_scenario_iphone}
    \end{subfigure}%
    \hspace{1em}%
    \begin{subfigure}[t]{.47\linewidth}
        \centering
        \includegraphics[width=\linewidth]{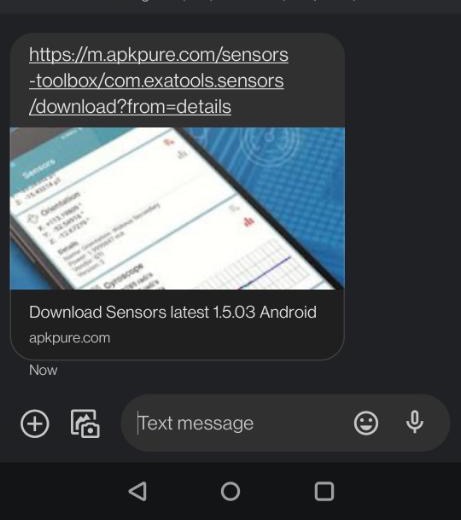}
        \caption{}
        \label{fig:attack_scenario_android}
    \end{subfigure}
\caption{Attack scenarios on different type of target devices (a) Apple headphone connection on iOS devices (b) malicious message on Android devices.}
\label{fig:attack_scenarios}
\vspace{-.5em}
\end{figure}

\subsection{Attack Vectors with Human Operation}
\label{sec:human_operates}
In the previous section, we presented the design of a static antenna array and
how it can be use to perform security oriented attacks on multiple devices in
several real scenarios. Although the antenna array is easy to build and use,
more powerful attacks can be carried out if the attacker has both access and the
ability to use a programmable mechanical system with our touch event injection
techniques, such as a miniature 3D printer~\cite{riscure} or robotic
arm\cite{keysight} commonly used in side channel analysis research. In this
case, our IEMI antenna more closely mimics the presence of a human finger and
the mechanical system mimics a human arm. To illustrate the capabilities of our
attack in this setting, we opt to manually maneuver our IEMI antennas to
simulate the attack with the mechanical system. With the short-tap,
press-and-hold and continuous omni-directional-swipe we achieve the following
security oriented attack outcomes. We believe these attacks are feasible and
practical to implement for a motivated attacker. 

\noindent\textbf{Send Message (Short-Tap)} With the short tap, we can send a
specific message to a recipient. In practice, such capabilities can be abused to
reply with confirmation messages when banks request text verification for
suspicious credit card transactions. In our experiment, we move our IEMI antenna
to generate short-tap touch events on top of the letters ``Y, E, S'' and the
enter position to send a confirmation message. The experiment is conducted on an
iPhone 11 Pro and a successful operation takes less than 10 seconds. 

\noindent\textbf{Send Money (Press-and-Hold)} A typical use case of
press-and-hold on iOS is providing shortcuts for certain functionalities with
minimum user interaction. For instance, Paypal allows iOS users to
hold-and-press the application icon to activate and send money by showing the QR
code without actually launching the application. We continuously apply our
interference signal on an iPad pro and point the IEMI antenna toward the Paypal
application to trigger this feature and evaluate the feasibility of such an
attack. We then move the antenna down to press on the "Send Money" option and
then turn off the interference signal to show the send money QR code. We
successfully launched this attack 7 out of 10 times at an attack distance
of 10mm. The completion time for every iteration of the attack was within 5
seconds. We found that human error, accidentally increasing the attack distance
while holding the antenna, was the reason for failed attack attempts.

\noindent\textbf{Unlock Gesture Lock Screen (Omni-Directional-Swipe)} A
significant achievement of our work compared to previous approaches is that we
can inject omni-directional-swipes with a controllable duration. As we show in
our video demonstration where we draw a figure with our IEMI antenna, if the
attacker can control the location of the IEMI antenna a gesture lock screen
unlock attack can be performed. We evaluate the feasibility by trying to unlock
a gesture lock protected application on an iPad Pro. The gesture lock we setup
has the shape of ``Z'' which includes 7 points at three different rows and
columns. This attack was successful 3 out of 5 times at an attack distance of
10mm. The completion time for every iteration of the attack was similarly within
5 seconds. The total travel distance of the IEMI antenna was 14 cm.

\section{Countermeasures} \label{sec:discussion}

\smallskip
\noindent \textbf{Force Detection:} Force and pressure add a new dimension on
top of existing touchscreen techniques. High end touchscreen
controllers~\cite{samsungA552} can detect the force applied on the touchscreen with a
scale from 1 to 10. The force sensors used in the touchscreen can detect subtle
differences in the amount of pressure of each touch. Since the introduced ghost
touches may not cause any pressure on the touchscreen, the underlying system can
check both force sensors and touchscreen controllers to filter out the ghost
touches. The test devices that we have do not have such features, so we use a barometer 
as a substitute for detecting the pressure on the touchscreen for those devices equipped with one. In our touch gesture detection application, we read the barometer value whenever a touch event occurs. For example, the barometer value on the Pixel 2 changes 0.3 hPa when the screen is pressed with a finger for more than 1 second. We successfully detect injected long press and swipes on a Pixel 2 using the barometer. However, this method is limited to Android devices with water resistance, otherwise the barometer value does not change even with a human finger pressing on the touchscreen.

\smallskip
\noindent \textbf{Low-Cost Accessory:} Apart from manufacture level
countermeasures, end users may use smartphone or tablet cases with metal front
covers to block all EM interference including the IEMI attacks. In fact, such
products are already available in the market~\cite{defendershield} and
originally designed to prevent the NFC card skimming
attack~\cite{francis2009potential}. To evaluate this countermeasure, we use a regular
phone case with front cover and tap the inner layer with Faraday Fabric. We keep the phone 
awake while using the phone with our customized phone case. Even though the thickness of the 
Faraday Fabric is only 0.28mm, it still defends our attack considerably well. We were no longer 
able to inject the touch events onto any test devices except for rare ghost touches at the edge of the touchscreen where the Faraday Fabric is not covered well. This countermeasure does not 
require any specific hardware or software to be present on the touchscreen device and can be implemented with minimum effort.

\section{Related Work} \label{sec:related}

\subsection{IEMI Attacks}

IEMI attacks have been applied to different devices and systems, including
medical devices~\cite{kune2013ghost}, smart
phones~\cite{kasmi2015iemi,esteves2018remote}, embedded
systems~\cite{selvaraj2018electromagnetic,kennedy2018susceptibility,delsing2006susceptibility},
autonomous vehicles\cite{richelli2019analog,dayanikli2020electromagnetic}, etc.

Among these attacks, Delsing \etal\cite{delsing2006susceptibility} examined the
effects of an IEMI attack on sensor networks and revealed the susceptibility of
sensor networks to high frequency (in GHz range) IEMI. Selvaraj
\etal\cite{selvaraj2018electromagnetic} further expanded this attack and
demonstrated that small circuits (i.e., embedded systems) are vulnerable to low
frequency IEMI with proper coupling. Kennedy \etal also studied how IEMI can be
used to create interference on the analog voltage input port of an Analog to
Digital Converter~\cite{kennedy2018susceptibility}.

Kune \etal conducted comprehensive analysis of IEMI attacks against analog
sensors and demonstrated IEMI attacks on cardiac medical devices by remotely
injecting forged signals~\cite{kune2013ghost} that cause pacing inhibition and
defibrillation. In this paper, the authors also demonstrated how to inject audio
signals on microphones remotely and proposed digital mitigations to verify and
clean the input signal. Kasmi and Esteves~\cite{kasmi2015iemi,esteves2018remote}
exploited the voice assistant on smart phones to perform remote inaudible
command injection attacks against smartphone headphone cables using fine tuned
EM signals. 

\subsection{Touchscreen Attacks} 

Various attacks targeting touchscreens have been presented in the past.
These attacks are primarily focused on passive information exfiltration,
e.g., displayed content, via different carriers including
microphone~\cite{genkin2019synesthesia}, EM~\cite{hayashi2014threat} or
mmWave signal~\cite{li2020wavespy}. In addition, only two
papers~\cite{maruyama2019tap, wangghosttouch} are published to perform
active touchscreen attack using IEMI. Maruyama \etal\cite{maruyama2019tap}
presented Tap'n Ghost, a new class of active attack against capacitive
touchscreens, which leverages an injected noise signal and programmed NFC
tag to force a victim mobile device to perform unintended operations.
However, this attack can only be conducted along with user touches due to
the skewed spatial distribution. On the contrary, our touchscreen IEMI
attack can cause intentional ghost touches on a capacitive touchscreen
without any user interaction. A recent touchscreen attack, Ghosttouch
~\cite{wangghosttouch}, similarly used EMFI to inject taps and row/column
based swipe gestures. Although the attack is more advanced than Tap'n
Ghost, it relies on detecting the correct driving signal from the
touchscreen and synchronizing it with IEMI signal to induce accurate touch
events. However, we find that the driving mechanism is significantly
different on different smartphones, which makes the attack less feasible in
a real attack scenario. As shown in Appendix
Figure~\ref{fig:scanning_signal_devices}, the measured driving signal from
five different touchscreen devices are entirely different. The Nexus 5X
smartphone used in Ghosttouch shows a clear synchronization pattern. On the
other hand, other smartphones use a parallel driving mechanism which is
difficult to synchronize with. Ghosttouch works well on sequential driving
based touchscreens. Unfortunately this is no longer a popular option for
the most recently released touchscreens. Furthermore, Ghosttouch is limited
to either column or row based swipe gestures due to the synchronization.
Our attack does not need to perform synchronization, nor rely on a specific
type of driving mechanism to inject stable short-tap, long-press, and
omni-directional-swipe touch events to realize practical attacks.

\section{Conclusions and Future Work} \label{sec:con}

In this paper, we first developed theory for a novel IEMI attack on modern
capacitive touchscreens to generate ghost touches. The theory was then validated
in both simulations and experimental demonstrations. We identify that such a
vulnerability exists in almost all capacitive touchscreen-based devices under
radiated IEMI attacks. The mechanism of the induced ghost touches cause is
analyzed based on the operating principle of touch sensing. The critical field
strength that can generate ghost touches is calculated, along with the critical
frequencies at which the touchscreens are more vulnerable to IEMI attacks. The
IEMI attack is successfully demonstrated on a series of commercial touchscreens
of laptop, smartphone, and tablets under various attack scenarios. We elaborate
on the features affecting our IEMI attack, including table material, table
thickness, phone locations, and antenna interference. Using our antenna array,
screen locator, and touch event detector, we design and evaluate the first
end-to-end touchscreen attack in real scenarios. We address several limitations
presented in previous touchscreen attacks. We further evaluate the proposed
countermeasures against our attack. 

In the future, we plan to increase our attack distance and attack accuracy
by using different antenna designs, i.e., longer waveguide (copper needle),
far-field phased array antenna, and Yagi-Uda (directional) antenna. We plan
to evaluate phased array antenna and Yagi-Uda antenna to programmatically
generate the focused E field from far so that we can address the current
table thickness limitation. On the other side, phased array antenna and
Yagi-Uda antenna can carry significant implementation challenges compared
to a copper needle antenna.

\section{Acknowledgment}\label{sec:ack}
We genuinely appreciate the reviewers for all their constructive suggestions.
This work is supported by National Institute of Standards and Technology, Intel 
and National Science Foundation under award number 1818500.

\clearpage

\bibliographystyle{IEEEtran}
\bibliography{bib/emi, bib/sslpub}

\appendix
\addcontentsline{toc}{section}{Appendices}
\renewcommand{\thesubsection}{\Alph{subsection}}
\setcounter{equation}{0}
\renewcommand{\theequation}{\thesubsection-\arabic{equation}}
\setcounter{figure}{0}
\renewcommand\thefigure{\thesubsection-\arabic{figure}}
\setcounter{table}{0}
\renewcommand\thetable{\thesubsection-\arabic{table}}  

\subsection{The scanning mechanism of touchscreens}
\label{app:scanning}

\begin{figure*}[!htb]
    \centering
    \includegraphics[width=\textwidth]{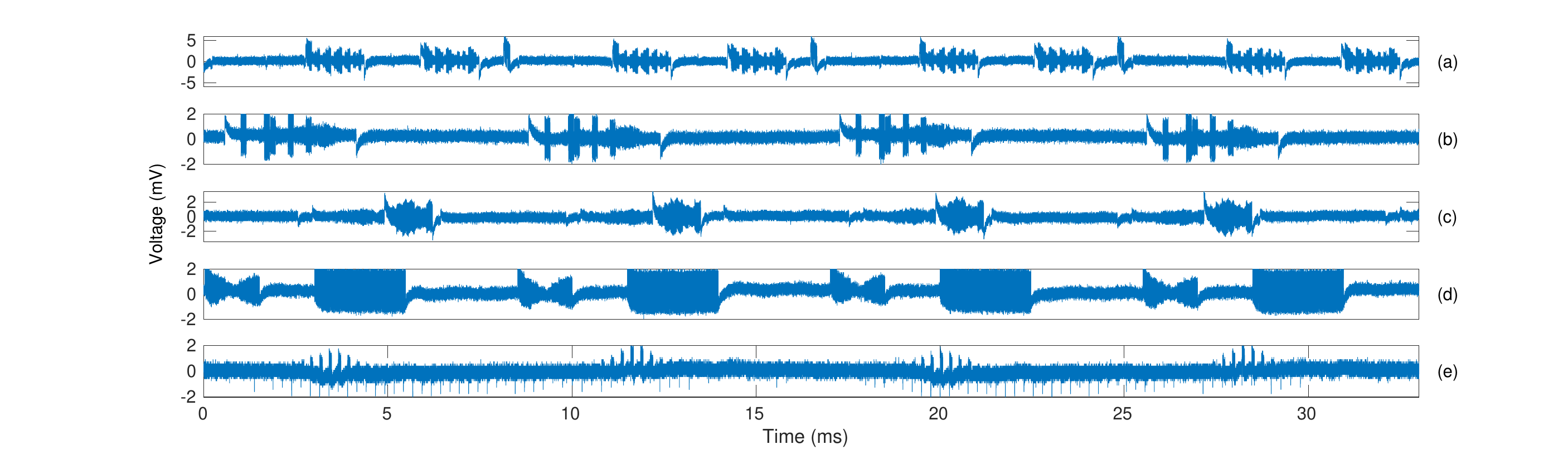}%
    \caption{Scanning signal of different touchscreen devices (a) iPad Pro 2020  (b) iPhone 11 Pro (c) Oneplus 7 Pro (d) Pixel 2 (e) Nexus 5X 
    \label{fig:scanning_signal_devices}}%
\end{figure*}

As we explained in Section~\ref{sec:phone_locator}, there are two type of
scanning mechanism mainly used by modern touchscreen, sequential driving method
and parallel driving method. As shown in Ghosttouch~\cite{wangghosttouch}, this
most recent touchscreen attack relies on the synchronization of sequential
driving signal to precisely inject touch events. However, such approach limits
the attack to sequential scanning type touchscreen. As illustrated in
Figure~\ref{fig:scanning_signal_devices}, the scanning signal from the test
devices we own are significantly different. We further find that latest
touchscreen devices commonly use parallel driving method instead, which makes
the synchronization based attack no longer feasible. Even with the sequential
driving method, different type of touchscreen can show significantly different
pattern. On the contrary, our attack does not reply on any particular scanning
method of touchscreen to work.

\subsection{Derivation of Equations of IEMI frequency}
\label{app:iemi_fre}

We assume that the electric field generated by the radiated IEMI is sinusoidal.
The noise current, $I_n$ in Fig. \ref{figs:equivalent_circuit_a}, is given as follows. 

\begin{equation}
    I_n=2\pi f_E C_M V_n\cos{\left(2\pi f_E \cdot t + \varphi_0\right)} \label{eqn:eqn19}
\end{equation}

\noindent
where $f_E$ is the E field frequency and $\varphi_0$ is the phase between $I_n$
and $S_2$ control signal in Fig.~\ref{figs:equivalent_circuit_b}. The waveforms
show the control signal of $S_2$ and the noise current caused by IEMI in
one period. The output voltage variation $V_{Tn}$ caused by the IEMI can then be calculated as follows. 

\begin{equation}
    V_{Tn}=-\frac{2\pi f_{E{C_MV}_n}}{C_s}\int_{0}^{T_s}{\cdot\cos{\left(2\pi f_E\cdot t+\varphi_0\right)}dt} \label{eqn:eqn20}
\end{equation}

\noindent 
where $T_s$ is the sensing time. Following (\ref{eqn:eqn20}), the $V_{Tn}$ at the end
of the sensing period can be calculated as follows. 

\begin{equation}
    V_{Tn}=-\frac{{C_MV}_n}{C_s}\left(sin{\left(2\pi f_E\cdot T_s+\varphi_0\right)}-sin\left(\varphi_0\right)\right) \label{eqn:eqn21}
\end{equation}

\noindent
During the IEMI injection period, $V_{Tn}$ is compared to the threshold
$V_{th}$. The control signal of the QT sensor is a periodical signal whose
frequency depends on the system clock frequency. More specifically, the sensing
time $T_s$ depends on the QT sensor switching frequency $f_{sw}$ and the duty
cycle $D_s$.

\begin{equation}
    T_s=\frac{D_s}{f_{sw}} \label{eqn:eqn22}
\end{equation}

\noindent
When we substitute (\ref{eqn:eqn22}) to (\ref{eqn:eqn21}), we have a more precise
way to compute the $V_{Tn}$ as shown in (\ref{eqn:eqn23}). 

\begin{equation}
    V_{Tn}=-\frac{{C_MV}_n}{C_s}\left(sin{\left(2\pi\cdot D_s\cdot\frac{f_E}{f_{sw}}+\varphi_0\right)}-sin\left(\varphi_0\right)\right) \label{eqn:eqn23}
\end{equation}

\noindent
From (\ref{eqn:eqn23}), it is clear that $V_{Tn}$ depends on the ratio of the IEMI signal frequency over the QT sensor operating frequency. The higher
$|V_{Tn}|$ is, the more significant the IEMI impact. Based on this
observation, we can conclude that the minimum interference occurs at
$f_{Emin}$, which can be calculated as follows. 

\begin{equation}
    f_{Emin}=\frac{kf_{sw}}{D_s}   \;\;\;\;\;  k=0,1,2,3,\dots \label{eqn:eqn24}
\end{equation}

\noindent
where $k$ is an integer. %
When $f_E=f_{Emin}$, $V_{Tn}$ in (\ref{eqn:eqn23}) is always zero, which
indicates that there is no interference. The maximum interference, on the other
hand, depends on the frequency of the IEMI signal as well as the phase shift
$\varphi_0$.

With the analysis in Section \ref{subsec:efield_strength}, we know that the output
voltage of QT sensor is usually compared with the threshold voltage every few
clock cycles. So combining (\ref{eqn:eqn8}) and (\ref{eqn:eqn23}),
the sum of output voltage variation of $M$ cycles, $V_{TnM}$, is given as follows.

\begin{equation}
    V_{TnM}=-\frac{C_M V_n}{C_s}\sum_{0}^{M}\left(sin{\left(2\pi f_E\cdot T_s+\varphi_M\right)}-sin\left(\varphi_M\right)\right) 
    \label{eqn:eqn25}
\end{equation}

\noindent where $\varphi_M$ can be calculated in (\ref{eqn:eqn26}).

\begin{equation}
    \varphi_M=\varphi_0+2\pi M\cdot\frac{f_E}{f_{sw}} 
    \label{eqn:eqn26}
\end{equation}

\noindent
Based on (\ref{eqn:eqn25}) and (\ref{eqn:eqn26}), we can calculate $f_E$ so that
the initial phase shift between $I_n$ and $S_2$ control signal remains constant
in each sensing duty cycle (see Fig. \ref{fig:equivalent_circuit} (b)). The
calculation of $f_E$ is shown below.

\begin{equation}
    f_E=nf_{sw} \;\;\;\;\; n=0,1,2,3,\dots  
    \label{eqn:eqn27}
\end{equation}

\subsection{Derivation of Equations of IEMI  Field Strength}
\label{app:efield}
A more detailed characterization of the E field interference is presented as follows. In Fig.~\ref{figs:E_field_interference_a}, $E_Z$ is the z component of the external E field, which generates voltage $V_n$ across the touch screen electrodes. $V_n$ can be calculated in (\ref{eqn:eqn9}).

\begin{equation}
    V_n=\int{E_Z\cdot d l}=E_Z\cdot d \label{eqn:eqn9}
\end{equation}

\noindent where $d$ is the distance between the electrodes. The charges ($Q_n$) caused
by the external E field can be derived as follows.

\begin{equation}
    Q_n=V_n\cdot C_M \label{eqn:eqn10}
\end{equation}

\noindent
where $C_M$ represents the mutual capacitance between the electrodes. It can be computed in (\ref{eqn:eqn11}).

\begin{equation}
    C_M=\varepsilon_0\varepsilon_r\frac{A}{d} \label{eqn:eqn11}
\end{equation}

\noindent where $\varepsilon_0$ is the permittivity of the free space and
$\varepsilon_r$ is the relative permittivity of the adhesive layer. $A$ is the
overlap area of the electrodes. From (\ref{eqn:eqn9}) – (\ref{eqn:eqn11}), we
can derive $E_Z$, the z component of the external E field.

\begin{equation}
    E_Z=\frac{Q_n}{\varepsilon_0\cdot\varepsilon_r\cdot A}=\frac{V_nC_M}{\varepsilon_0\cdot\varepsilon_r\cdot A} \label{eqn:eqn12}
\end{equation}

\noindent
Based on superposition theory, the voltage $V_{cN}$ which is added to the
input of the integrator in Fig.~\ref{figs:E_field_interference_b} can be
computed as follows.

\begin{equation}
    V_{cN}=V_c+V_n \label{eqn:eqn13}
\end{equation}

\noindent where $V_c$ is the voltage of $C_M$ due to $V_{in}$. The
output voltage, $V_{oN}$, under the external E field's interference is, therefore, as follows.

\begin{equation}
    V_{oN}=-\frac{C_M}{C_s}\left(V_c+V_n\right)=V_o+V_{Tn} \label{eqn:eqn14}
\end{equation}

\noindent

\end{document}